\documentclass[superscriptaddress,aps,twocolumn,showpacs,preprintnumbers,amsmath,amssymb]{revtex4-1}
\usepackage{graphicx}
\usepackage{dcolumn}
\usepackage{epstopdf}
\usepackage{xcolor}
\usepackage{color}
\usepackage{amssymb}
\usepackage{bm}
\usepackage{hyperref}
\usepackage{booktabs}
\usepackage[normalem]{ulem}

\graphicspath{
 {figure/}
}

\newcommand{ \be }{\begin{eqnarray}}
\newcommand{ \ee }{\end{eqnarray}}
\newcommand{ \la }{\langle}
\newcommand{ \lla }{\left\langle}
\newcommand{ \ra }{\rangle}
\newcommand{ \rra }{\right\rangle}
\def \mean#1 {{\la #1 \ra}}

\newcommand{ \pp}{pp}
\newcommand{ \pA}{p--A}

\newcommand{ \PbPb}{Pb--Pb}

\newcommand{ \dAu}{d--Au}
\newcommand{ \AonA}{A--A}

\newcommand{ \pt }{\textit{p}_{\rm T}}
\newcommand{ \gevc }{GeV/\textit{c}}

\newcommand{ \pti}{\textit{p}_{\rm T,i}}

\newcommand{ \dpt }{\rm{d}\textit{p}_{\rm T}}

\newcommand{ \DeltaPtPt}{\langle  \Delta \pt \Delta \pt \rangle}

\newcommand{ \dphi}{\rm{d}\varphi}

\newcommand{ \dphiOne}{\rm{d}\varphi_1}
\newcommand{ \dphiTwo}{\rm{d}\varphi_2}
\newcommand{ \Dphi }{\Delta \varphi}

\newcommand{ \deta }{\rm{d}\eta}
\newcommand{ \detaOne }{\rm{d}\eta_1}
\newcommand{ \detaTwo }{\rm{d}\eta_2}
\newcommand{ \Deta }{\Delta \eta}
\newcommand{ \etaPhi }{\left( \eta,\varphi \right)}

\newcommand{ \DetaDphi }{\left(  \Delta \eta, \Delta \varphi \right)}
\newcommand{ \rhoOne}{\rho_1}

\definecolor{dgreen}{cmyk}{1.,0.,1.,0.2}        
\definecolor{orange}{cmyk}{0.,0.353,1.,0.}    

\def \snn {\sqrt{\textit{s}_{_{\rm NN}}}}

\long\def\/*#1*/{}
\begin{document}

\title{\texorpdfstring{Differential two-particle number and momentum correlations  with the AMPT, UrQMD, and EPOS models in Pb-Pb collisions at $\sqrt{\textit{s}_{_{\rm NN}}}$= 2.76 TeV}{}}

\author{Sumit Basu}
\email{sumit.basu@cern.ch}
\affiliation{Lund University, Department of Physics, Division of Particle Physics, Box 118, SE-221 00, Lund, Sweden}
\affiliation{Department of Physics and Astronomy, Wayne State University, Detroit, 48201 USA}
\author{Victor Gonzalez}
\email{victor.gonzalez@cern.ch}
\affiliation{Department of Physics and Astronomy, Wayne State University, Detroit, 48201 USA}
\author{Jinjin Pan}
\affiliation{{Cyclotron Institute, Texas A\&M University, College Station, Texas 77843, USA}}
\author{Anders Knospe}
\affiliation{University of Houston, USA}
\author{Ana Marin}
\affiliation{GSI Helmholtzzentrum f\"ur Schwerionenforschung, Research Division and ExtreMe Matter Institute EMMI, Darmstadt, Germany}
\author{Christina Markert}
\affiliation{University of Texas at Austin, USA}
\author{Claude Pruneau}
\email{claude.pruneau@wayne.edu}
\affiliation{Department of Physics and Astronomy, Wayne State University, Detroit, 48201 USA}


\begin{abstract}
We report studies of charge-independent (CI) and charge-dependent (CD) two-particle differential number correlation functions, $R_2 \DetaDphi$, and transverse momentum correlation functions, $P_2 \DetaDphi$, of charged particles produced in \PbPb\ collisions at the LHC centre-of-mass energy $\sqrt{s_{\rm NN}} =$  2.76 TeV with the UrQMD, AMPT and EPOS models. 
Model calculations for  $R_2$ and $P_2$ correlation functions are 
presented for inclusive charged hadrons ($h^\pm$) in selected  transverse momentum ranges and with full azimuthal coverage in the pseudorapidity range $|\eta|< 1.0$.
We compare these calculations for the strength, shape, and particularly the width of
the  correlation functions with recent measurements of these observables by the ALICE collaboration. 
Our analysis indicates that comparative studies of $R_2$ and $P_2$  correlation functions
provide valuable insight towards the understanding of particle production in \PbPb\ collisions. We find, in particular,
that these models  produce quantitatively different magnitudes and  shapes for these correlation functions and none 
reproduce the results reported by the ALICE collaboration.
 Accounting for quantum number conservation in models, particularly charge conservation,  is mandatory to reproduce the detailed measurements
of number and transverse momentum correlation functions. 
\end{abstract}
\keywords{Correlation functions, QGP, Collectivity, Heavy Ion Collisions}
\pacs{25.75.Gz, 25.75.Ld, 24.60.Ky, 24.60.-k}
\maketitle
\section{Introduction}
\label{sec:introduction}

 Integral and differential  correlation functions measurements are essential tools
for the study of  proton-proton (\pp) and heavy-ion (\AonA) collisions at relativistic energies. 
Azimuthal correlations functions have provided evidence for the 
existence of anisotropic flow in  \AonA\ collisions~\cite{Adams:2005dq,Adcox:2004mh,Aamodt:2010pa,Aamodt:2011by,CMS:2013bza,CMS-harmonic,ATLAS:2012at,Atlas_wei}, (approximate) quark scaling of flow 
 coefficients in  \AonA\ collisions at  RHIC and LHC~\cite{PhysRevLett.98.162301,Abelev:2014pua,Adam:2016nfo,ATLAS:2012at}, 
 as well as evidence for the presence of long range correlations in smaller systems such as \pp\ and \pA\ collisions~\cite{Esumi:2017xvf,MalgorzataALICE:2017Corr,X.Zhu:2013JetCorrelation,Abelev:2014mda,Bernardes:2017eup,Sirunyan:2018toe}. Differential two-particle (number) correlation 
 functions have also enabled the discovery of jet quenching  at RHIC~\cite{Adler:2002tq,PhenixA.AdarePRC:2008JetQuenching} and its detailed 
 characterization  in  \AonA\ collisions at both RHIC and LHC~\cite{Chatrchyan:2011sx}. Many other correlation 
functions, including number and transverse momentum correlation 
functions~\cite{Adams:2005ka,Agakishiev:2011fs} have been studied  at RHIC and 
LHC to better understand the particle production dynamics and elucidate  
the properties of the matter produced in \pp\ and  \AonA\ collisions~\cite{Bass:2000az,S.PrattPRL:2000BalFun1st,Pratt:2015jsa,Aggarwal:2010ya,Adams:2003kg,Abelev:2013csa}. Among these, the recent measurements~\cite{PhysRevC.100.044903} of number correlation, 
$R_2$, and differential 
transverse momentum correlation, $P_2$, defined in 
Sec.~\ref{sec:definition}, have enabled 
independent confirmation of the collective nature of the 
azimuthal correlations observed in \PbPb\ 
collisions~\cite{Adam:2017ucq}, as well as 
the identification of noticeable differences in the $\Deta$ and 
$\Delta\varphi$ dependence of these correlation functions. Indeed, 
measurements by the ALICE collaboration~\cite{Adam:2017ucq,PhysRevC.100.044903} show that the near-side 
peak of both charge independent (CI) and charge dependent  (CD) correlations is  significantly narrower, at any given  \AonA\ collision centrality, in $P_2$ than in $R_2$ correlation functions in both longitudinal and azimuthal directions. This confirms~\cite{M.SharmaPRC:2009Methods} that comparative measurements of $P_2$ and $R_2$ correlation functions may provide additional sensitivity to the underlying particle production mechanisms involved in 
heavy-ion collisions.  

In this work, we compare calculations of the $R_2$ and $P_2$ correlation functions with the UrQMD~\cite{Bass:1998ca, Bleicher:1999xi, Petersen:2008dd,Petersen:2008kb}, AMPT~\cite{Lin:2004en}, and EPOS~\cite{DRESCHER200193,PhysRevC.82.044904,PhysRevC.89.064903} models with the measurements recently reported by the ALICE collaboration~\cite{Adam:2017ucq,PhysRevC.100.044903}, with a particular focus  on charged particles 
produced in the range $0.2 < \pt \le 2.0$ \gevc. 
We seek  to establish, in particular, whether the three selected models can reproduce the distinctive features of CD and CI combinations of these correlation functions. For instance, in \PbPb\ collisions at $\sqrt{s_{\rm NN}}=$ 2.76 TeV, the  correlators 
feature  near- and away-side structures featuring specific dependence on collision centrality. The ALICE collaboration reported
that the near-side of the $P_2$ correlator is typically much narrower than that of its $R_2$ counterpart~\cite{Adam:2017ucq,PhysRevC.100.044903}. Additionally, the width of the
near-side peak of CD correlation functions is observed to narrow considerably from peripheral to central collisions while the CI correlation 
functions exhibit  broadening and a significant change of shape in more central collisions. 

This paper is organized as follows. Section~\ref{sec:definition} 
presents definitions of  the $R_2$ and $P_2$ correlation functions studied in this 
work and describes how they are computed, whereas sec.~\ref{sec:mechanisms} presents a discussion of the particle production and transport properties these correlation functions are sensitive to.  The UrQMD, AMPT, and EPOS models, and the conditions under which they were used to generate \PbPb\ events, are briefly introduced in Sec.~\ref{sec:models}. Correlation functions obtained with  the three models  are presented  in Sec.~\ref{sec:Results} and conclusions are summarized in  Sec.~\ref{sec:summary}.

\section{Correlation functions definition} 
\label{sec:definition}

The $R_2$ and $P_2$  correlation functions (hereafter also called correlators) are defined in 
terms of single and two particle densities expressed as functions 
of the particles pseudorapidity $\eta$ and azimuthal angle $\varphi$ 
\be
\rho_1^{\alpha} \etaPhi &=&\frac{1}{\sigma} \frac{\rm{d}^2\sigma^{\alpha}}{\deta \dphi}, \\ 
\rho_2^{\alpha\beta}(\eta_1, \varphi_1,\eta_2, \varphi_2)&=& \frac{1}{\sigma} 
\frac{\rm{d}^4\sigma^{\alpha\beta}}{\detaOne\dphiOne\detaTwo \dphiTwo},  
\ee
where  $\sigma$  represents the inelastic  cross-section,  $\sigma^{\alpha}$ is the single particle production cross-section of particles of type $\alpha$, and $\sigma^{\alpha\beta}$ is the pair production of particle types $\alpha$ and $\beta$. In the context of this paper, we limit the discussion to correlation function of charged particles. The indices $\alpha$ and $\beta$ thus stand for positive $(+)$ and negative $(-)$ particles. 

The   $R_2$ correlator is defined as a two-particle cumulant  normalized by the product of single 
particle densities according to 
\begin{equation}
R_2^{\alpha\beta} (\eta_1, \varphi_1, \eta_2, \varphi_2) = \frac{\rho_2^{\alpha\beta}(\eta_1, \varphi_1, \eta_2, \varphi_2) }{ \rho_1^{\alpha}(\eta_1, \varphi_1)\rho_1^{\beta}(\eta_2, \varphi_2)}-1,
\end{equation}
whereas the $P_2$ correlator is defined in 
terms of the momentum correlator $\DeltaPtPt$ 
normalized by the square of inclusive mean transverse momentum, 
$\lla \pt \rra$, to make it dimensionless  
\begin{equation}
P_2^{\alpha\beta} (\eta_1, \varphi_1,\eta_2, \varphi_2)=\frac{\DeltaPtPt^{\alpha\beta}(\eta_1, \varphi_1,\eta_2, \varphi_2)}
{\la \pt\ra^2}.
\end{equation}
The $\la \Delta \pt\Delta \pt \ra^{\alpha\beta}$ differential correlator~\cite{M.SharmaPRC:2009Methods}  is defined according to 
\begin{widetext}
\begin{equation}
\la \Delta \pt\Delta \pt \ra^{\alpha\beta}(\eta_1, \varphi_1,\eta_2, \varphi_2) =
\frac{\int_{p_{\rm T,\min}}^{p_{\rm T,\max}} {\rm d} p_{\rm T,1} \,{\rm d} p_{\rm T,2} \, \rho_2^{\alpha\beta}(\vec p_{1},\vec p_{2}) \Delta p_{\rm T,1} 
\Delta 
p_{\rm T,2} } 
{\int_{p_{\rm T,\min}}^{p_{\rm T,\max}} {\rm d} p_{\rm T,1}\, {\rm d} p_{\rm T,2} \,
\rho_2^{\alpha\beta}(\vec p_{1},\vec p_{2}) }
\end{equation}
\end{widetext}
where $\Delta\pti =  \pti - \la \pt \ra$ and $\la \pt \ra$ is the 
inclusive mean transverse momentum  defined  according to  $\la 
\pt \ra = \int_{p_{\rm T,\min}}^{p_{\rm T,\max}}\rhoOne \pt  \dpt 
/\int_{p_{\rm T,\min}}^{p_{\rm T,\max}}\rhoOne  \dpt$.

The correlators $R_2$ and $P_2$  are  reported  
as  functions of the differences $\Deta= \eta_1 - \eta_2$ and 
$\Dphi=\varphi_1 - \varphi_2$ by averaging across the mean 
pseudo-rapidity 
$\bar \eta = \frac{1}{2}(\eta_1 + \eta_2)$ and the mean
azimuthal angle 
$\bar \varphi = \frac{1}{2}(\varphi_1 + \varphi_2)$ acceptance 
according to 
\begin{widetext}
\begin{equation}
O(\Deta, \Dphi) 
= \frac{1}{\Omega(\Deta)} \int O(\eta_1, \varphi_1,\eta_2, 
\varphi_2) \delta(\Dphi - \varphi_1 + \varphi_2)  \dphiOne \dphiTwo  \\ 
\times  \delta(\Deta - \eta_1 + \eta_2)  \detaOne \detaTwo,
\end{equation}
\end{widetext}
where $O$ represents either of the $R_2$ or $P_2$ correlators and $\Omega(\Deta)$ is the width of the acceptance in $\bar{\eta}$  at a given value of $\Deta$. The angle difference $\Dphi$ is  calculated modulo $2\pi$ and shifted by $-\pi/2$ for convenience of representation in the figures. The analysis is carried out for charge combination pairs $(\alpha\beta)=(+-)$, $(-+)$, $(++)$, and $(--)$ separately. Like-sign pairs correlations are averaged to yield LS correlators, ${\rm LS} =\frac{1}{2}[ (++) + (--)]$, and US  correlators are obtained  by averaging  $(+-)$ and $(-+)$ correlations, ${\rm US} = \frac{1}{2}[ (+-) + (-+)]$. The LS and US correlators are then combined to yield charge-independent (CI) and charge-dependent (CD) correlators defined according to  
\be
\label{eq:CI}
O^{\rm (CI)} &=& \frac{1}{2}\left[O^{\rm (US)} + O^{\rm (LS)}\right], \\ 
\label{eq:CD}
O^{\rm (CD)} &=& \frac{1}{2}\left[O^{\rm (US)} - O^{\rm (LS)}\right],
\ee 
respectively. The CI correlator measures the 
average of all correlations between charged particles while the CD 
correlator is sensitive to the difference between US and LS pairs and is as such determined largely by charge conservation. 

The $R_2$(CD) correlator is strictly proportional to the 
charge balance function (BF)~\cite{S.PrattPRL:2000BalFun1st} when the 
yields of positive and negative 
particles are equal~\cite{C.PruneauPRC:2002Fluct}.

\begin{widetext}
\begin{equation}
\begin{aligned}
B^{\alpha\beta}(\vec{p}_{\alpha},\vec{p}_{\beta}) = \frac{1}{2} 
 \left\{ 
 \rho_1^{\beta^-} 
 \left[
 R_2^{\alpha^+\beta^-}(\vec{p}_{\alpha^+},\vec{p}_{\beta^-}) - {R_2^{\alpha^-\beta^-}(\vec{p}_{\alpha^-},\vec{p}_{\beta^-})} 
 \right] +
\rho_1^{\beta^+}
\left[ 
R_2^{\alpha^-\beta^+}(\vec{p}_{\alpha^-},\vec{p}_{\beta^+}) - {R_2^{\alpha^+\beta^+}(\vec{p}_{\alpha^+},\vec{p}_{\beta^+})} 
\right] 
\right\}
\end{aligned}
\label{eq:BFR2}
\end{equation}
\end{widetext}
where labels $\alpha$ and $\beta$ represent the types (species) of particles considered. Balance functions are sensitive to mechanisms of particle production and transport in A--A collisions. They were first considered to investigate the presence of delayed hadronization~\cite{Bass:2000az,Pratt:2002BFLH}, but they were recently also shown to be particularly sensitive to the hadro-chemistry of the collision systems as well as the diffusivity of light quarks~\cite{Pratt:2015PC,Pratt:2019pnd}. The AMPT, UrQMD, and EPOS models are already known to successfully reproduce the $p_{\rm T}$ spectrum of produced particles, i.e., the single particle densities $\rho_1^{\alpha}$ obtained with  measurements of (identified and inclusive) charged particles. Given the balance function is proportional to those yields but its shape and structure are primarily determined  by the normalized cumulants ${R_2^{\alpha\beta}}^{\rm (CD)}$, we  limit our discussion to  a comparison of the calculated correlators ${R_2^{\alpha\beta}}^{\rm (CD)}$ to those reported by the ALICE collaboration. 

\section{Properties of the $R_2$ and $P_2$ correlators.} 
\label{sec:mechanisms}

Heavy ion collisions are rather complex phenomena involving diverse particle production and transport mechanisms.  It is thus of interest to briefly consider what physics insight can be brought about by the  $R_2$ and $P_2$ correlators.

At very large collision energy,   the yields of anti-particles and particles are nearly equal \cite{Aamodt:2010dx,Abelev:2013vea} and so are, essentially, correlators of same sign
particles, i.e., measured correlators for $(+,+)$ and $(-,-)$ pairs are essentially indistinguishable. But  conservation laws, including (electric) charge conservation, baryon number conservation, strangeness conservation, as well as energy-momentum conservation  significantly constrain the particle production. Interesting insight may thus be  provided by comparing same- (LS) and opposite-sign (US) particle pairs, e.g., $h^+h^+$ and $h^+h^-$, or baryon-baryon and baryon-anti-baryon particle pair correlations. It 
is of interest, in particular, to consider what correlation  features the  LS and US pairs may have in common and identify those that distinguish them. This is readily accomplished with the  study of charge independent (CI) and charge dependent (CD)  combinations of the LS and US correlation functions, defined by Eqs.~(\ref{eq:CI},\ref{eq:CD}). CI combinations of the $R_2$ and $P_2$ correlators reveal correlation features that are common to both LS and US pairs while the CD combinations emphasize their differences. 

Prior studies have shown that  CI and CD combinations of the differential correlation functions $R_2$ and $P_2$ are sensitive to several mechanisms of particle production and transport in pp, p--A, and A--A collisions~\cite{PruneauWWND2017,PhysRevC.100.024909,Adam:2017ucq}. Among others, these include energy-momentum conservation, quantum number conservation, response to pressure gradients and different levels of opacity, resonance decays, as well as jet production and quenching, etc. A full discussion of the sensitivity of the $R_2$ and $P_2$ correlators and their CI and CD combinations to all 
these facets is beyond the scope of this work but   Tab.~\ref{tab:CorPhys} provides a brief synopsis of their properties and response to these different facets of heavy-ion collisions.
\begin{table*}[htb]
\begin{tabular}{|l|c|c|}
\hline
Concerned Physics Processes  &$R_2^{\rm CD}$,$P_2^{\rm CD}$ & $R_2^{\rm CI}$,$P_2^{\rm CI}$ \\
\hline 
1. Coulomb + HBT &\checkmark & \checkmark \\
2. Jet cross-section, fragmentation, quenching, angular ordering & \checkmark  & \checkmark \\
3. Energy-momentum conservation   &\checkmark&  \\
4. Quantum number (Q,S,B) conservation& \checkmark&\checkmark \\
5. Anisotropic flow &   \checkmark       & \checkmark \\
6. Resonance decays &\checkmark&\checkmark \\
7. String/Color tube fragmentation and other long range correlations & \checkmark &\checkmark\\
8. Transport - Radial flow & \checkmark &\checkmark \\
9. Quark diffusivity& \checkmark &  \\
10. Two-stage hadronization & \checkmark & \\ 
\hline 
\end{tabular}
\caption{Sensitivity of the $R_2^{\rm CD}$, $P_2^{\rm CD}$, $R_2^{\rm CI}$, and $P_2^{\rm CI}$ correlators to the different physics processes of relevance for particle production in heavy-ion collisions.}
\label{tab:CorPhys}
\end{table*}

Given the $P_2$ observable has so far received only a limited amount of attention, it is interesting to discuss its properties in some detail. The $P_2$ correlator features an explicit dependence on the momenta of the particles relative to the mean transverse momentum, $\la p_{\rm T}\ra$. One can then expect that correlation structures observed with $P_2$ should be qualitatively different than those observed with $R_2$. Specifically, by virtue of the dependence on $\Delta p_{\rm T}\Delta p_{\rm T}$, $P_2$ is sensitive to the
``hardness" of the correlated particles. On the one hand, if correlations are dominated by a preponderance of particle pairs   with $p_{\rm T}> \la p_{\rm T}\ra$ or
$p_{\rm T}< \la p_{\rm T}\ra$, then $P_2$ is expected to be positive definite. On the other hand, if correlations are dominated by pairs featuring one particle with $p_{\rm T}>\la p_{\rm T}\ra$ and the other with $p_{\rm T}< \la p_{\rm T}\ra$, then $P_2$ is expected to feature negative values on average. Furthermore, a change from positive to negative values is expected as a function of  $\Delta\eta$ and $\Delta\varphi$  in the vicinity of
the near-side peak for correlations involving jet fragments as a specific $p_{\rm T}$ vs. $\theta$ ordering ($\theta$ being the angle of particle emission 
relative to the initial parton direction) as shown in Ref.~\cite{PhysRevC.100.024909}. Such change  from positive to negative values might also be observed in
the presence of resonance decays with large radial boost~\cite{PruneauWWND2017}. Either way, the presence of such a shift  from positive to negative values 
vs. $\Delta\eta$ and $\Delta\varphi$ is expected to lead to a narrower near-side peak in $P_2$ correlations than in $R_2$ correlations. 
The width difference, however, should be sensitive to the details of the jet angular ordering and/or the relative magnitude of resonance decay contributions to these correlators. Two-prong  decays of resonances at rest would, nominally,  yield back-to-back two-particle correlation structures. In practice, thermal and strong radial flow fields produced in A--A collisions kinematically focus progeny particles into  a relatively narrow near-side peak surrounding  $\Delta\eta =0 $, $\Delta\varphi=0$. 
Moreover, the fragmentation of jets is known to yield a somewhat narrow correlation peak  in $\Delta\eta$ vs. $\Delta\varphi$ coordinates, while back-to-back jet production leads to a relatively broad away-side correlation structure centered at $\Delta\varphi = \pi$ and typically extending over a wide range of pseudo-rapidity differences in these correlators. 
The  strength and shape of the near-side correlation peaks of $R_2$ and $P_2$ are thus sensitive to the relative abundances of hadronic resonances as well as the radial flow profile that accelerates them. Moreover, although the correlators $R_2$ and $P_2$ nominally measure the same pairs and thus the same correlations, the explicit dependence of $P_2$ on the product of deviates $\Delta p_{\rm T} \Delta p_{\rm T}$ provides sensitivity to the $p_{\rm T}$ ordering of the particles~\cite{PhysRevC.100.024909}. 

A joint study of the differential correlators $R_2$ and $P_2$ thus provide sensitivity to the details of the hadronic cocktail, that is, the hadro-chemistry of the system, as well as its transport characteristics.   Furthermore, initial spatial anisotropy, particularly in heavy A--A systems, is known to generate considerable pressure gradients that drive anisotropic particle production in the transverse plane of  these collisions. Such anisotropies, characterized by flow coefficients $v_n$, $n\ge 2$,  are found to extend over a very wide range of rapidity differences at RHIC and LHC energies. Recent ALICE measurements and comparison of $P_2$  and $R_2$ correlators in fact provided further support to the notion that azimuthal (i.e., $\Delta\varphi$) modulations find their origin in the initial spatial anisotropy and geometry of colliding nuclei~\cite{Adam:2017ucq}. A comparison of the long range behavior of $R_2$ and $P_2$ correlators thus also yield sensitivity to flow and non-flow contributions.

It is also worth noting that the integral of the $P_2$ correlator is sensitive to  event-by-event fluctuations of  the average $p_{T}$ of particles, and by extension, to event-by-event  fluctuations of the system temperature, $\Delta T^{2}$, a quantity of interest  towards the determination of the heat capacity of the medium~\cite{Sumit}. Finally, also note that CD combinations of $P_2$ and $R_2$   correlators (of US and LS pairs) should have, for the same reasons, much additional sensitivity to the presence of charge balancing pairs (i.e., pairs of negative and positive particles produced by a common charge conserving process). Differences between the $P_2$ and $R_2$ correlators are thus expected  to exhibit good sensitivity to the details of the particle production.
Based on the above discussion, one concludes that the  correlators $R_2$ and $P_2$ together provide sensitivity to a broad range of A--A collisions essential features, including  the hadro-chemistry of the collisions as well as transport properties of the medium. As such, they provide useful tools to test the performance  of  proton--proton and heavy-ion collision models.  Authors of this work have already reported on $R_2$ and $P_2$ correlation functions obtained with PYTHIA~\cite{P.SkandsJHEP:2006Pythia6.4Manual,P.SkandsPRD:2010PythiaTune}
and HERWIG~\cite{G.CorcellaJHEP:2001Herwig6.5Manual} and found these two models qualitatively reproduce many of the correlation features observed experimentally~\cite{PhysRevC.100.024909}. Interestingly, however, they ``predict" correlation functions that  quantitatively differ from one another. Measured $R_2$ and $P_2$ correlation functions thus provide new discriminant tools to test the performance and adequacy of these models. Turning our attention to heavy-ion collisions, it stands to reason that the discriminant character of these correlators can also provide a tool to challenge the performance of heavy-ion models. Specifically, given the three models considered in this work simulate particle production using distinct approaches, it is of interest to find out whether they can reproduce the strength, width, and shape of near-side correlation peaks, the presence of a $\Delta\eta$  extended away-side correlation ridge, as well as the strong elliptic and triangular $\Delta\varphi$ modulations observed experimentally in \PbPb\ collisions~\cite{PhysRevC.100.044903}.

\section{Monte Carlo Models} 
\label{sec:models}

We compare and contrast the $R_2$ and $P_2$ ALICE measurements in \PbPb\ collisions at $\snn = 2.76$ TeV~\cite{Adam:2017ucq,PhysRevC.100.044903}  with calculations with most up to date versions of  the AMPT, UrQMD, and EPOS models. These latest versions feature model parameters  tuned to reproduce  measured single particle spectra, relative yields, as well as flow parameters,  in contrast to earlier versions used before first  results  were reported by LHC experiments~\cite{Abreu:2007kv}. Recent versions of the three models have had considerable success in describing features of measured data at RHIC and the LHC but have also encountered limitations~\cite{PhysRevLett.105.022302}. 

The Ultra-relativistic Quantum Molecular Dynamics model (UrQMD)~\cite{Mitrovski:2008hb} is a microscopic many-body transport model initially designed to study hadron-hadron, hadron-nucleus and  heavy-ion collisions from $E_{\rm Lab} = 100$ A$\cdot$MeV to $\snn =$ 200 GeV. 
It was enhanced to include an intermediate hydrodynamical stage (hybrid configuration) \cite{Petersen:2008dd} to describe  the hot and dense medium produced in heavy-ion collisions  at top RHIC and LHC energies.  UrQMD describes the early stages of collisions in terms of partonic interactions and string formation, whereas its hadronic transport component, which describes the later stages of system evolution includes a full spectrum of hadrons, including 55 baryon mass states (up to 2.25 GeV/$c^2$) and 32 meson mass states and their respective anti-particles. All isospin-projected states to
elementary cross sections are used to fit to available proton--proton, proton--neutron or pion--proton data and the isospin symmetry is used whenever possible to obtain a complete description of scattering cross sections. UrQMD additionally uses additive quark model assumptions to account for otherwise unknown cross section such as those of hyperon-baryon resonance scatterings.
The model additionally guarantees that   quantum numbers are conserved globally
event-by-event on the Cooper-Frye hypersurface.
The original and  hybrid versions of the model have proven successful in describing  features of datasets acquired   at  SPS, RHIC, and LHC energies~\cite{Petersen:2011sb,Abbas:2013bpa}, including $p_{\rm T}$ spectra, average $p_{\rm T}$ values, as well anisotropic flow coefficients. The model thus appear to successfully describe the underlying radial flow field of particles in the final state. And, given the hadronic transport component of the model includes a full complement of baryon and meson resonances, one would expect it should adequately reproduce contributions to $R_2$ and $P_2$ arising from resonance decays.  It is less clear, however, how earlier stages of the collisions (partonic level) might manifest themselves in these correlators. Comparison of $R_2$ and $P_2$ correlators computed with UrQMD shall then provide a rather comprehensive assessment of the development and evolution of partonic level correlations and their manifestations in the hadronic final state. 

Our analysis is based on~$\sim$ 340K minimum bias events generated with the hybrid configuration of UrQMD Version 3.4. 
The program was compiled with the LHC option. The equation of state used during the hydrodynamical evolution includes a crossover deconfinement phase transition. The particle distributions are generated according to the Cooper–Frye prescription from the iso-energy density hypersurface, which is constructed using the Cornelius hypersurface finder. A cell size of 0.1 fm is used in the fluid description, that expands over 2 units of rapidity. The transition time to hydrodynamics is  at 0.5 fm.

AMPT~\cite{Lin:2004en} is a multi-phase transport consisting of 
several components of pre-existing codes such as the Heavy Ion Jet Interaction Generator (HIJING) for generating the initial conditions, Zhang's Parton Cascade (ZPC) for modeling partonic scatterings, the Lund string fragmentation model or a quark coalescence model for hadronization, and A Relativistic Transport (ART) model for treating hadronic scatterings. It has had relative success in reproducing several observables measured in heavy-ion collisions at both RHIC and LHC energies, including single-particle transverse momentum spectra of light particles~\cite{Abbas:2013bpa,Basu:2016dmo}  and the strength of transverse anisotropy harmonics~\cite{Lin:2004en,Solanki:2012ne}.  However, it  has encountered mitigated success in the prediction of correlation and fluctuation observables~\cite{PhysRevLett.105.022302}. AMPT can be operated in different modes (rescattering on/off, string-melting on/off) but our analysis is here limited to   rescattering-on (RON) and string-melting-on events (SON) known to be more apt at producing large resonance excitations and stronger radial flow profiles.  Given AMPT also includes a full complement of hadronic resonances and reproduces transverse momentum spectra and anisotropic coefficients rather well, one would expect it should also be able to describe the long range behavior of $R_2$ and $P_2$ correlators as well as their near-side peaks.  Comparison of correlation functions computed with AMPT with ALICE data shall thus also provide an important test of its ability to properly describe the underlying correlation strengths and the details of the radial flow profile of produced particles.   

A total of~$\sim$ 200K RON/SON minimum bias events were generated and used towards the production of the correlation functions presented in this work. Note, however, that the  version ampt-v1.26t7-v2.26t7 used in this work is known to violate charge conservation in specific cases. We thus do not expect it should properly describe the detailed shape and strength of CD  correlators but it might nonetheless be successful in describing the general features of CI correlation functions as well as salient features  of the CD correlation functions.

The EPOS model implements a multiple scattering approach based on partons and Pomerons (parton ladders), with special emphasis on high parton-densities~\cite{DRESCHER200193,PhysRevC.82.044904,PhysRevC.89.064903,Werner:2007bf}.
In its latest version~\cite{Anders}, EPOS3 also implements a prescription to distinguish core and corona zones of particle production within the colliding nuclei. The low-density region, i.e., the corona, is treated using  Regge theory  to compute the particle production, whereas the high-density region, known as  the core,  is described with  hydrodynamic equations of motion. A   Cooper-Frye prescription is used to implement the production 
of hadrons by the core component. 
This core/corona model has had considerable success in reproducing features of  \pp\ and \dAu\ collisions. With the addition of this core corona distinction, the model has also had good success
in reproducing the centrality evolution of resonance and strange particle production in heavy-ion collision systems~\cite{Anders}. It also reproduces anisotropic flow features reported by 
many experiments. While the core component of the model does not  properly handle charge conservation on an event-by-event basis and is thus not expected to reproduce features of CD correlations, we seek to find out whether it can reproduce the main features of CI correlation functions
as well as the main features of CD correlators. 

A total of~$\sim$ 320K minimum-bias \PbPb\ EPOS3 events, generated on the University of Texas Stampede  supercomputer  and requiring in excess of 100,000 CPU hours, were processed in this analysis. Model parameters used for the generation of events analyzed in this work are identical to those used in~\cite{Anders} (UrQMD on). Herein,  we refer to the EPOS3 model as EPOS for the sake of simplicity.

While our selection of the UrQMD, AMPT, and EPOS models for a comparative study with ALICE measurements of the $R_2$ and $P_2$ correlators was in part driven by practical considerations, it is important to recognize that they feature representative and comprehensive efforts, by the theoretical community, to model the many aspects and components of A--A collisions. Features, success, and concerns of these models are succinctly summarized in Tab.~\ref{EventGen_Comparision}. One notes that while the models have some common features, they are also based, broadly speaking, on rather different underlying approaches. And yet, all three  models have had considerable success in the description of many observables reported at RHIC and the LHC. It is thus clear that  the set of observables used so far to test the underlying physics of these models is not sufficiently discriminating to falsify the models. One can wonder, however, whether ``new" observables such as $R_2$ and $P_2$ might provide additional discriminating power to determine which of the underlying model components are correct or essential and which should be, perhaps, discarded unless they can be tuned, in a near future, to reproduce the added constraints provided by measurements of $R_2$, $P_2$ and other related correlation functions.   Given all three models are rather complex and multi-stage components of heavy-ion collisions, it is somewhat difficult, ab initio, to exactly identify how the contributions of their different components shall determine the strength and shape of the $R_2$ and $P_2$ correlators. Table~\ref{EventGen_Comparision} provides a brief survey of the respective features of these models that should influence the shape and form of the charge dependent and charge independent $R_2$ and $P_2$ correlators.
The sensitivity of these correlators to specific  physics processes has been already summarized in Tab.~\ref{tab:CorPhys}~\cite{Basu:2020ldt}.
\begin{table*}[htb]
\begin{tabular}{|l|l|l|l|}
\hline
\textbf{Models} &
  \multicolumn{1}{c|}{\textbf{UrQMD}} &
  \multicolumn{1}{c|}{\textbf{AMPT}} &
  \multicolumn{1}{c|}{\textbf{EPOS}} \\ \hline
Main Features &
  \begin{tabular}[c]{@{}l@{}}Ideal Hydro + \\ Hadronic cascade\end{tabular} &
  Microscopic transport model &
  \begin{tabular}[c]{@{}l@{}}Soft (QGP or Hydro like) +\\ Hard (QCD) component\end{tabular} \\ \hline
\begin{tabular}[c]{@{}l@{}}  Correlation \\     expected\end{tabular} &
  \begin{tabular}[c]{@{}l@{}}Resonance decays\\  + hadronic phase\end{tabular} &
  \begin{tabular}[c]{@{}l@{}}String fragmentation + \\ Zhang Parton Cascade \\ and Quark Coalescence\end{tabular} &
  Hard process + afterburner \\ \hline
\begin{tabular}[c]{@{}l@{}}  Anisotropic\\     flow\end{tabular} &
  \begin{tabular}[c]{@{}l@{}}Fluid cell momentum \\ anisotropy + hadronic \\ afterburner\end{tabular} &
  \begin{tabular}[c]{@{}l@{}}Escape mechanism + \\ A relativistic transport\end{tabular} &
  Soft process + afterburner \\ \hline
  Success &
  \begin{tabular}[c]{@{}l@{}}Particle productions,\\ pseudorapidity distribution,\\ multiplicity density spectra,\\ flow, \end{tabular} &
  \begin{tabular}[c]{@{}l@{}}Particle productions,\\ pseudorapidity distribution,\\ multiplicity density spectra,\\ flow, nuclear modification\end{tabular} &
  \begin{tabular}[c]{@{}l@{}}Particle productions,\\ pseudorapidity distribution,\\ multiplicity density spectra,\\ flow, nuclear modification\end{tabular} \\ \hline
  Concerns &
  \begin{tabular}[c]{@{}l@{}}Cooper-Frye could dilute\\  the correlations, \end{tabular} &
  \begin{tabular}[c]{@{}l@{}}No medium interaction, \\ partial charge conservation\end{tabular} &
  \begin{tabular}[c]{@{}l@{}}For Soft (core) part Cooper-Frye \\ could dilute the Correlations, Only\\ hard (corona) could show charge \\ correlations\end{tabular} \\ \hline
\end{tabular}
\caption{Summary of characteristics, successes, and concerns associated with the UrQMD, AMPT, and EPOS models, and their potential ability to properly model $R_2$ and $P_2$ correlators.
}
\label{EventGen_Comparision}
\end{table*}

\section{Model Calculations} 
\label{sec:Results}

\begin{figure*}[htb] 
\begin{center} 
\includegraphics[scale=0.30,clip=true,trim=101pt 5pt 91pt 1pt]{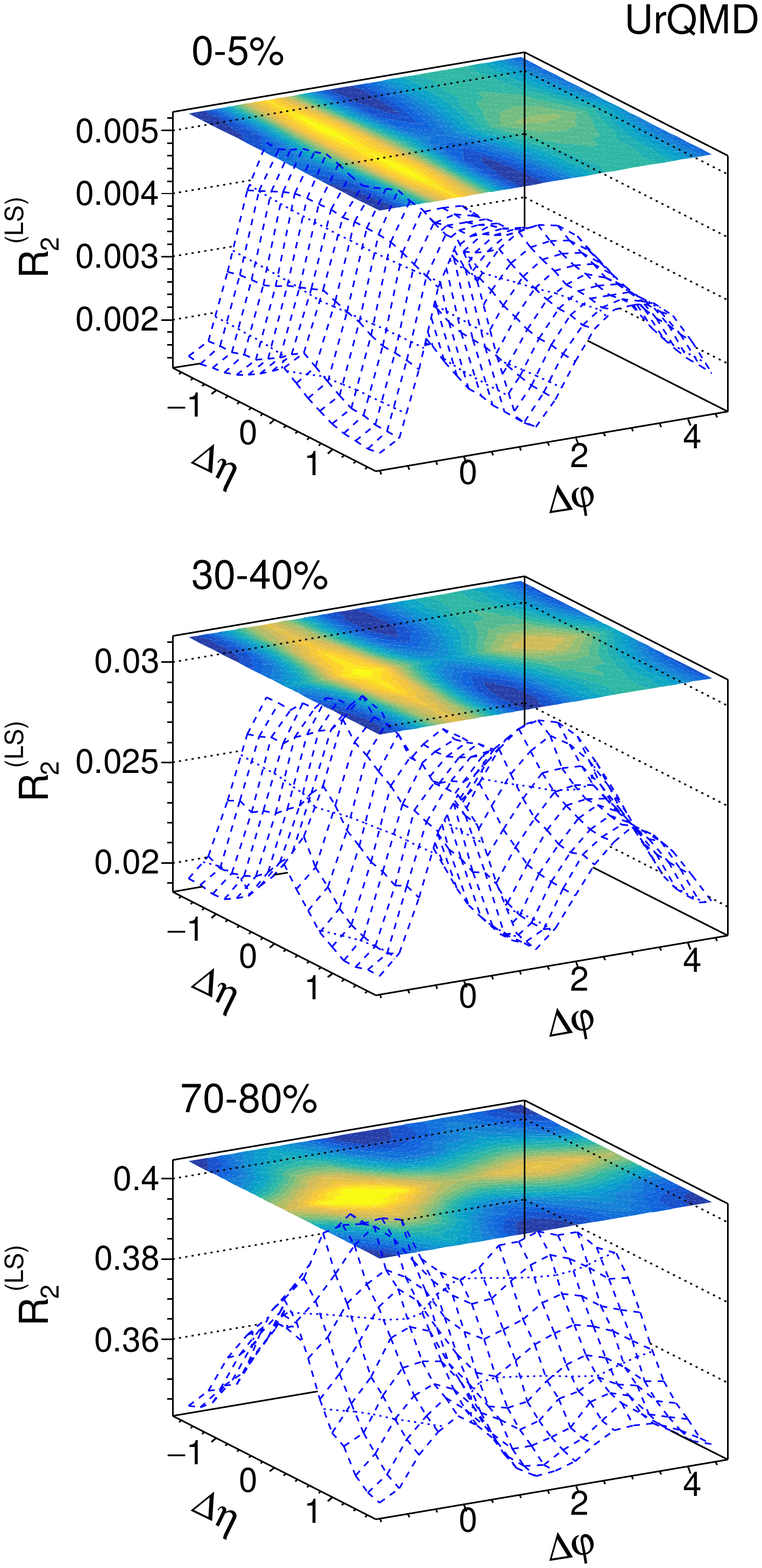}
\includegraphics[scale=0.30,clip=true,trim=101pt 5pt 91pt 1pt]{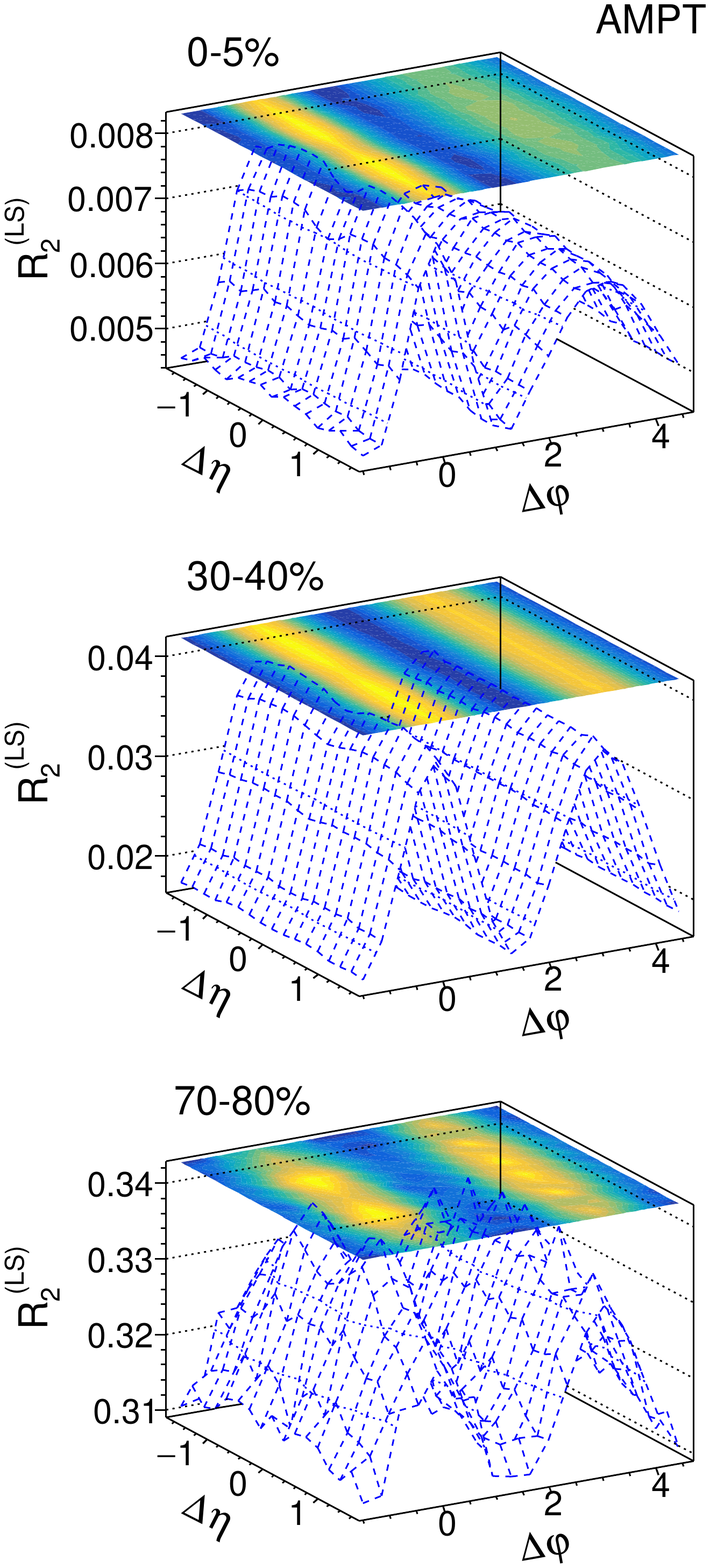}
\includegraphics[scale=0.30,clip=true,trim=101pt 5pt 91pt 1pt]{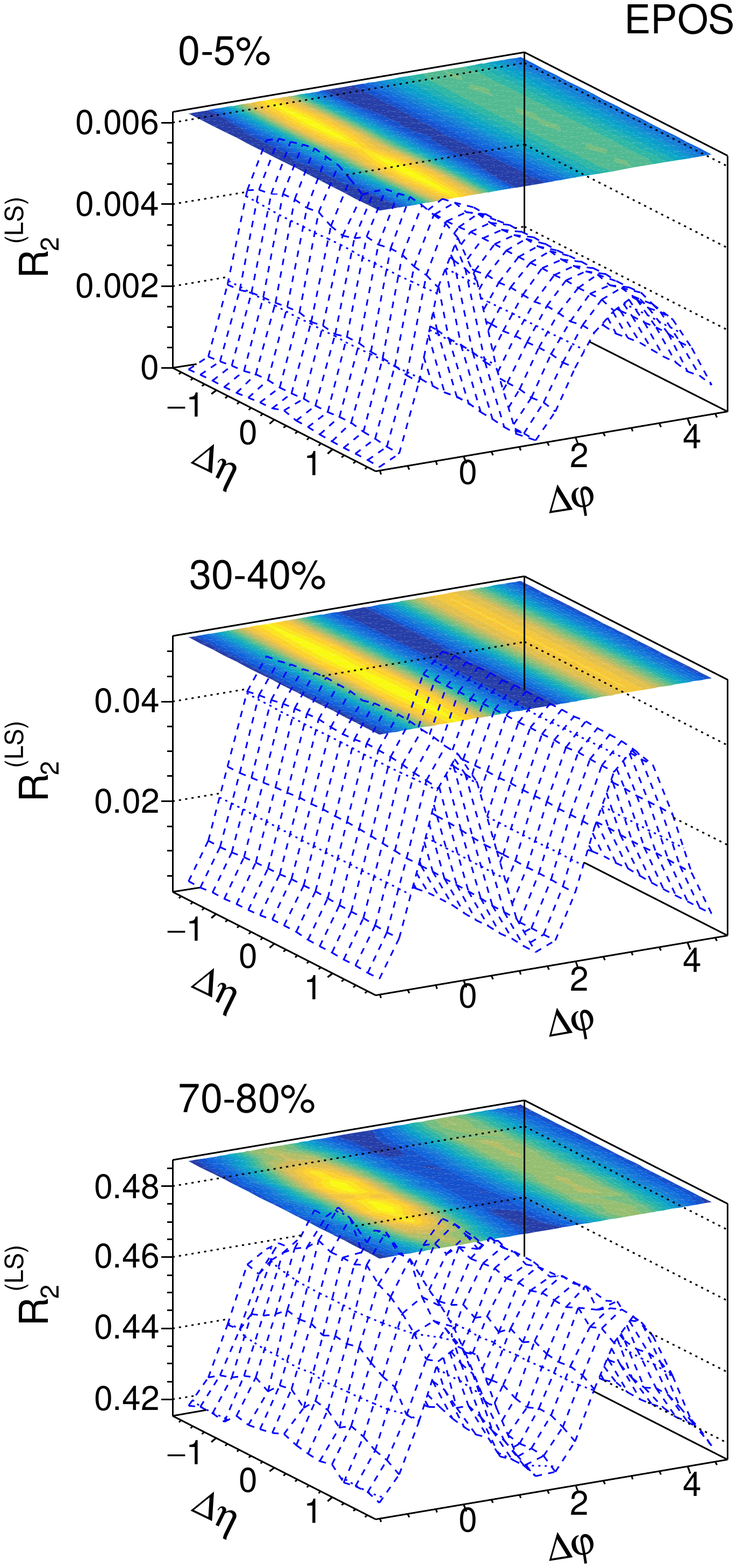}
\includegraphics[scale=0.30,clip=true,trim=75pt 5pt 91pt 1pt]{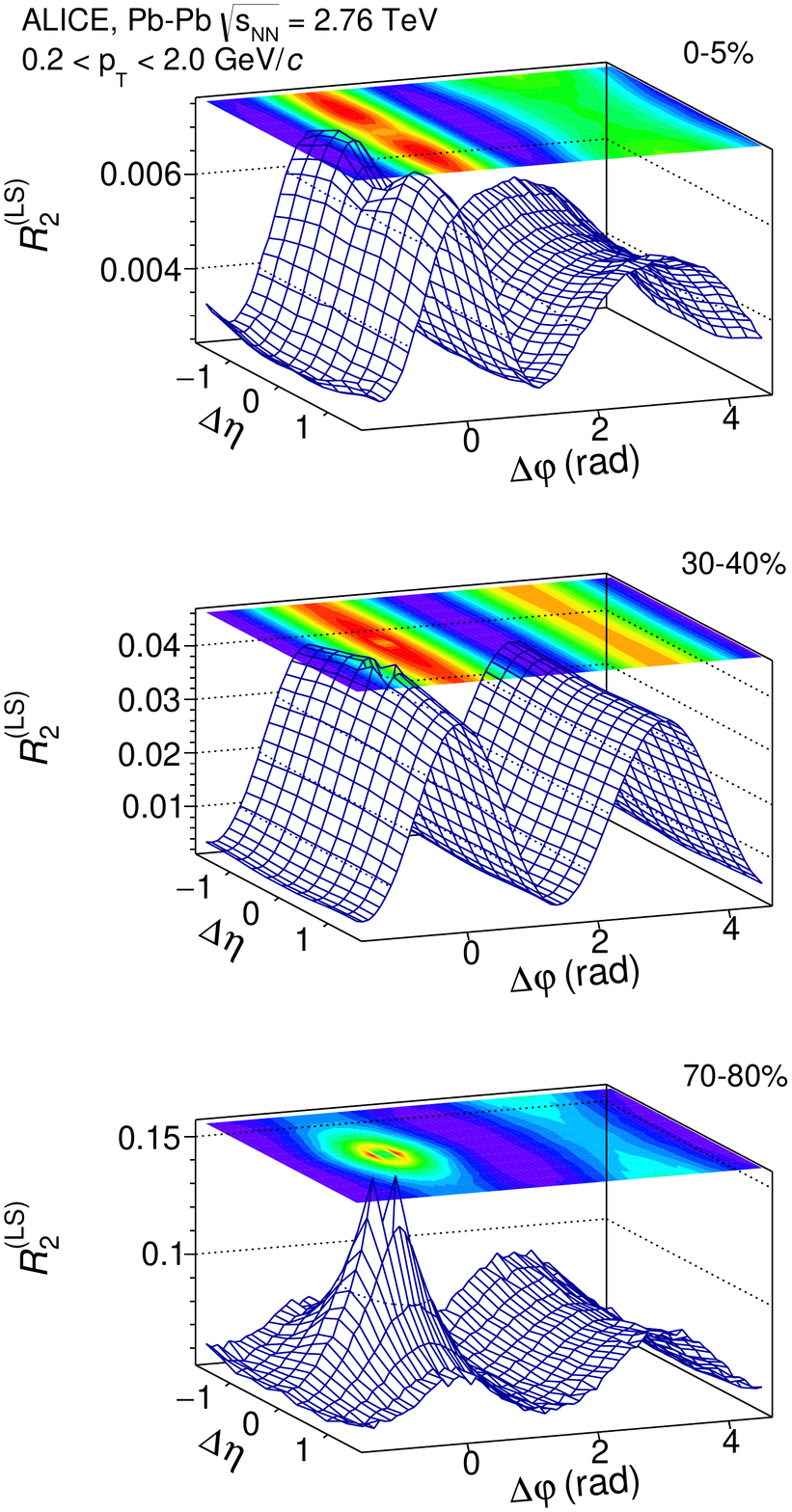}
\caption{Correlators $R_2^{(LS)}$ obtained with the UrQMD, AMPT (SON/RON) and EPOS models  compared to  correlators measured by the ALICE collaboration~\cite{PhysRevC.100.044903}  in \PbPb\  collisions at $\snn = 2.76$  TeV for three representative collision centrality ranges. Correlators are based on charged hadrons in the range $0.2 <  \pt \leq 2.0$~\gevc. See text for details.}
\label{fig:cR2LS} 
\end{center} 
\end{figure*}

\begin{figure*}[htb!] 
\begin{center} 
\includegraphics[scale=0.30,clip=true,trim=101pt 5pt 91pt 1pt]{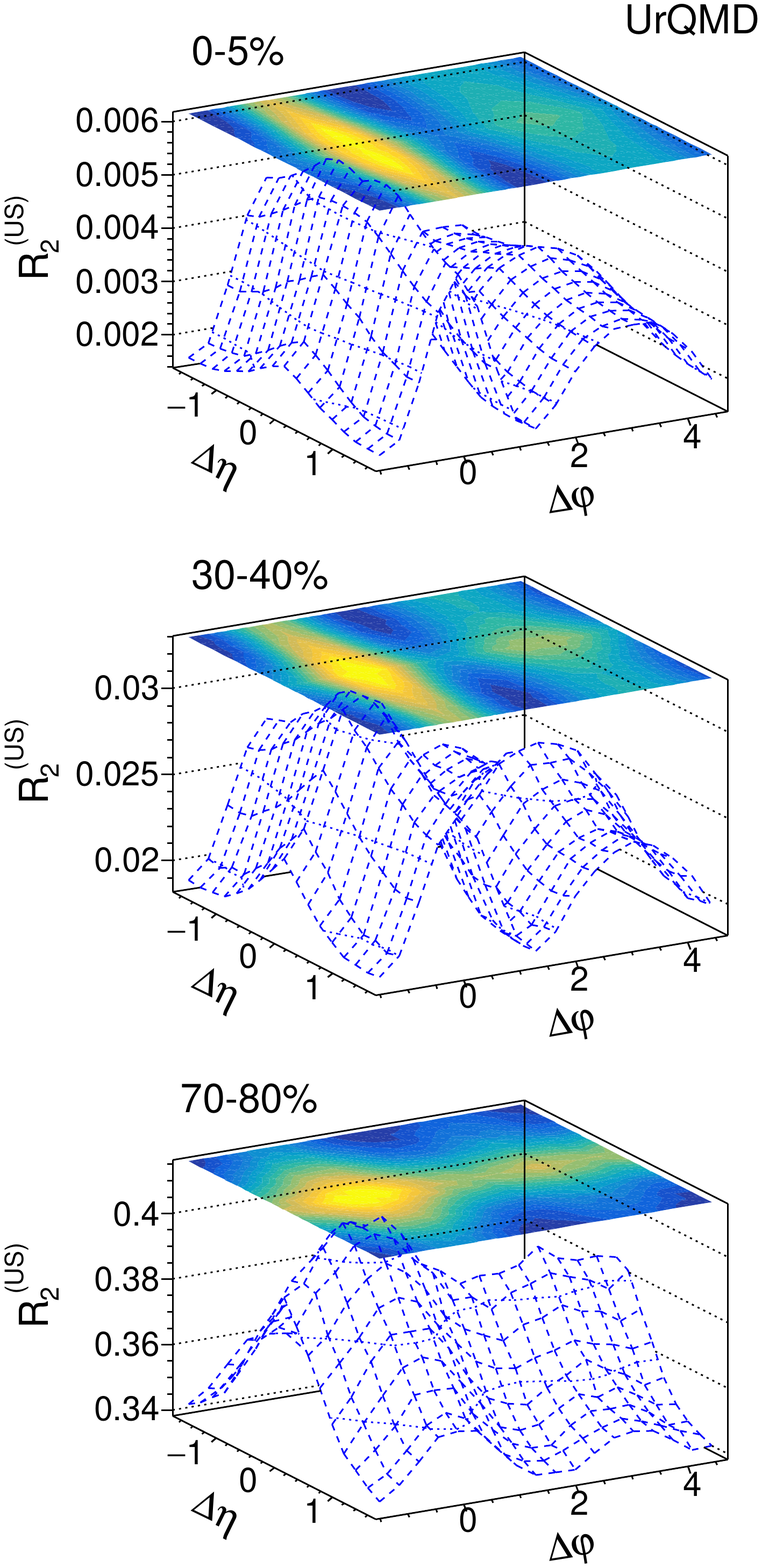}
\includegraphics[scale=0.30,clip=true,trim=101pt 5pt 91pt 1pt]{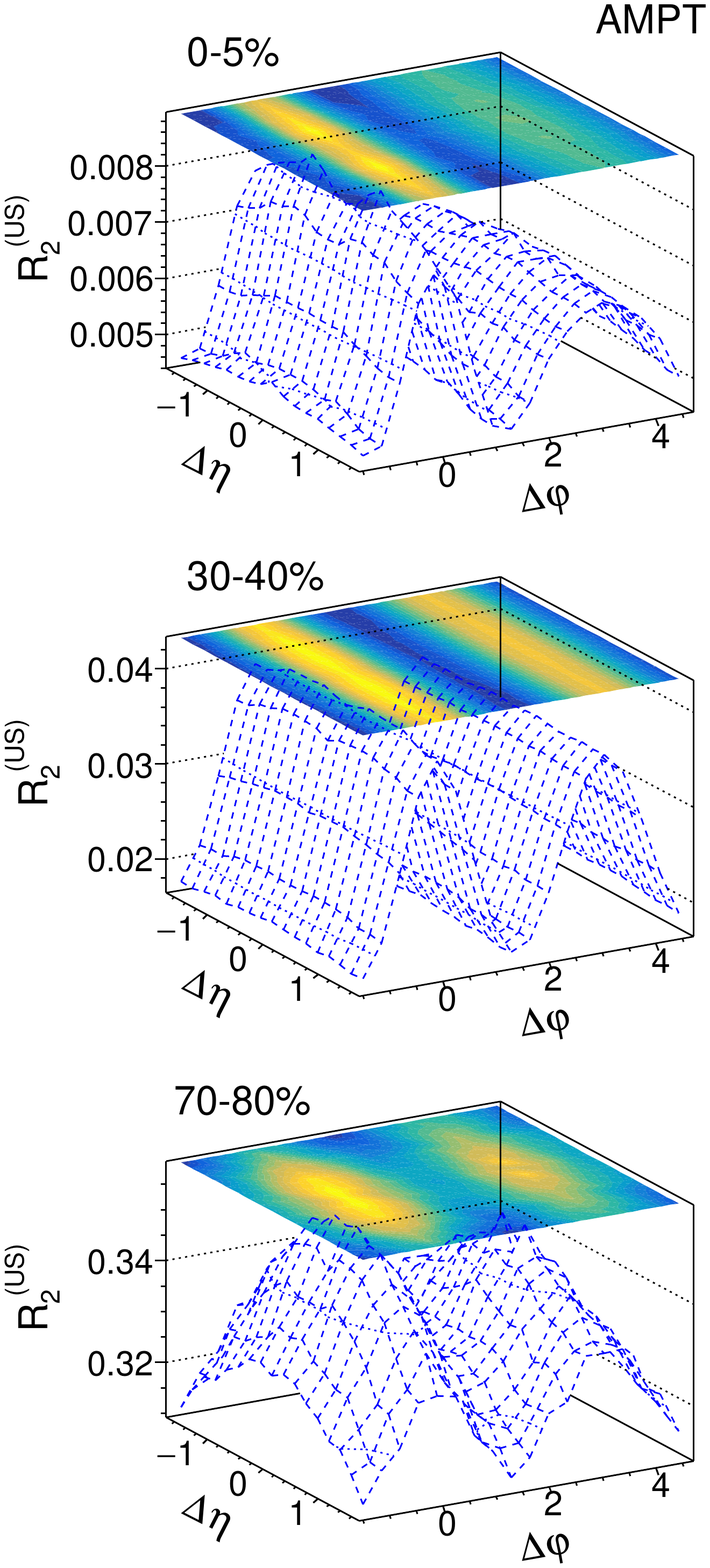}
\includegraphics[scale=0.30,clip=true,trim=101pt 5pt 91pt 1pt]{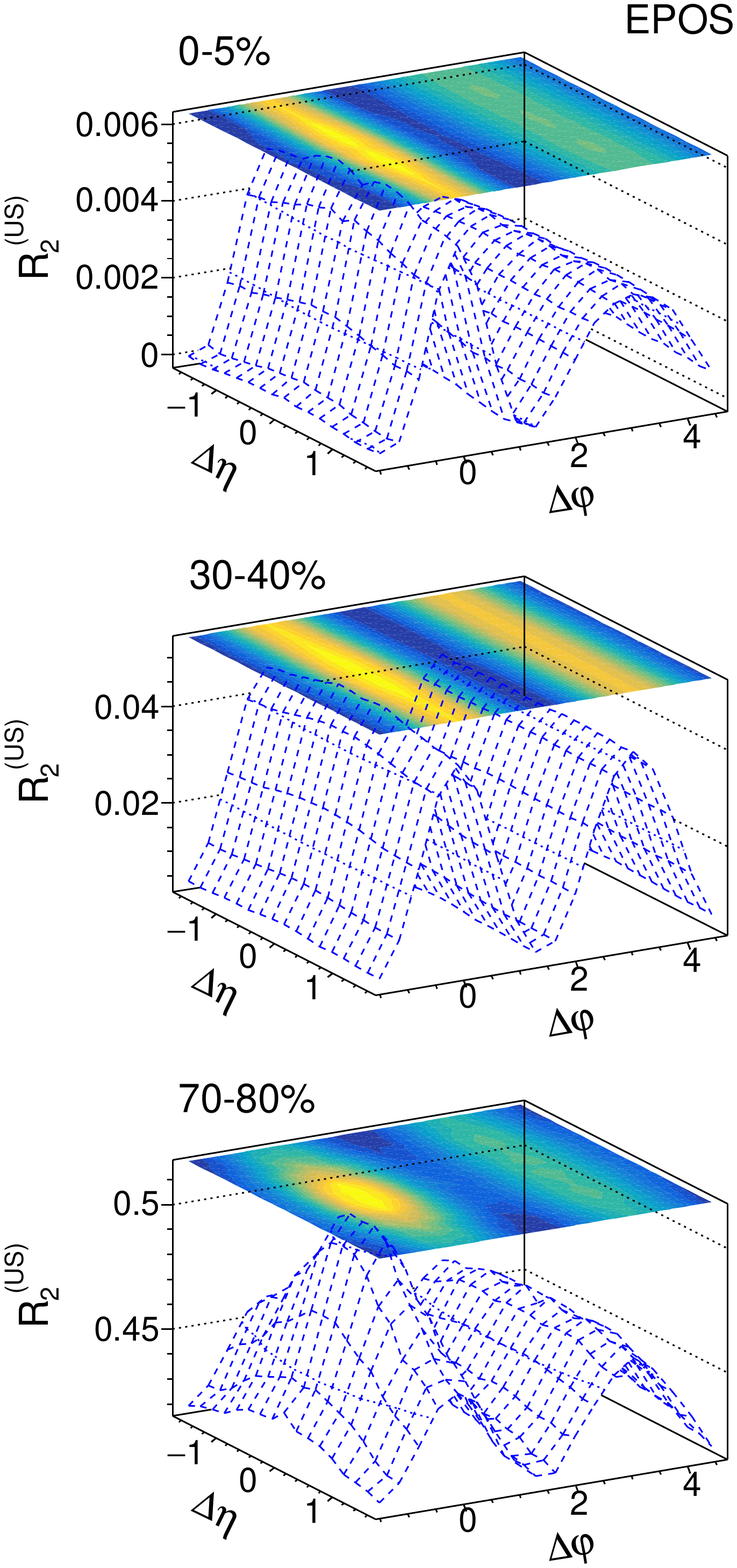}
\includegraphics[scale=0.30,clip=true,trim=75pt 5pt 91pt 1pt]{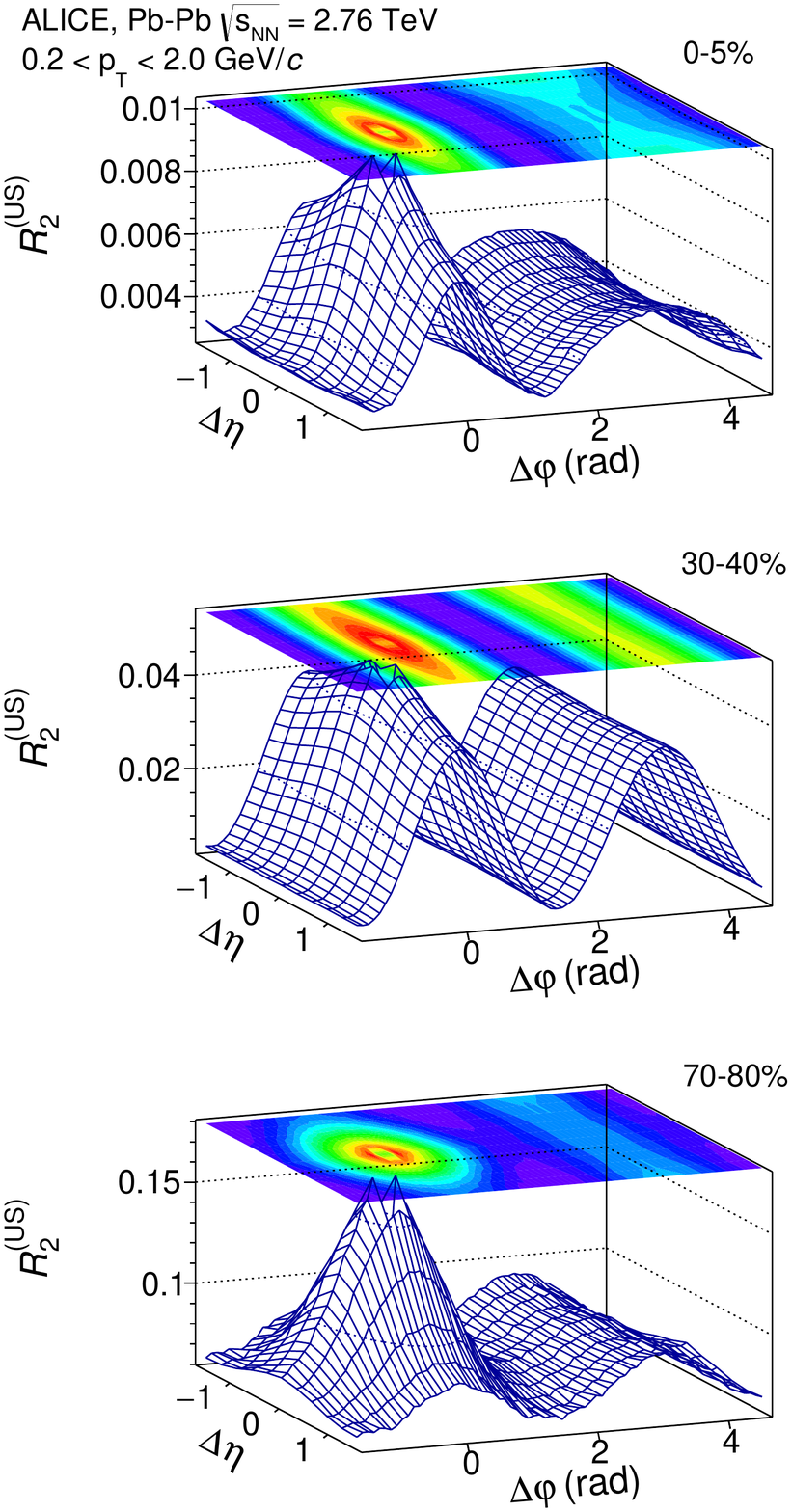}
\caption{Correlators $R_2^{(US)}$ obtained with the UrQMD, AMPT (SON/RON) and EPOS models compared to  correlators measured by the ALICE collaboration~\cite{PhysRevC.100.044903} in \PbPb\  collisions at $\snn = 2.76$  TeV for three representative collision centrality ranges. Correlators are based on charged hadrons in the range $0.2 <  \pt \leq 2.0$~\gevc. See text for details.}
\label{fig:cR2US}
\end{center} 
\end{figure*}

\begin{figure*}[htb!] 
\begin{center} 
\includegraphics[scale=0.30,clip=true,trim=101pt 5pt 91pt 1pt]{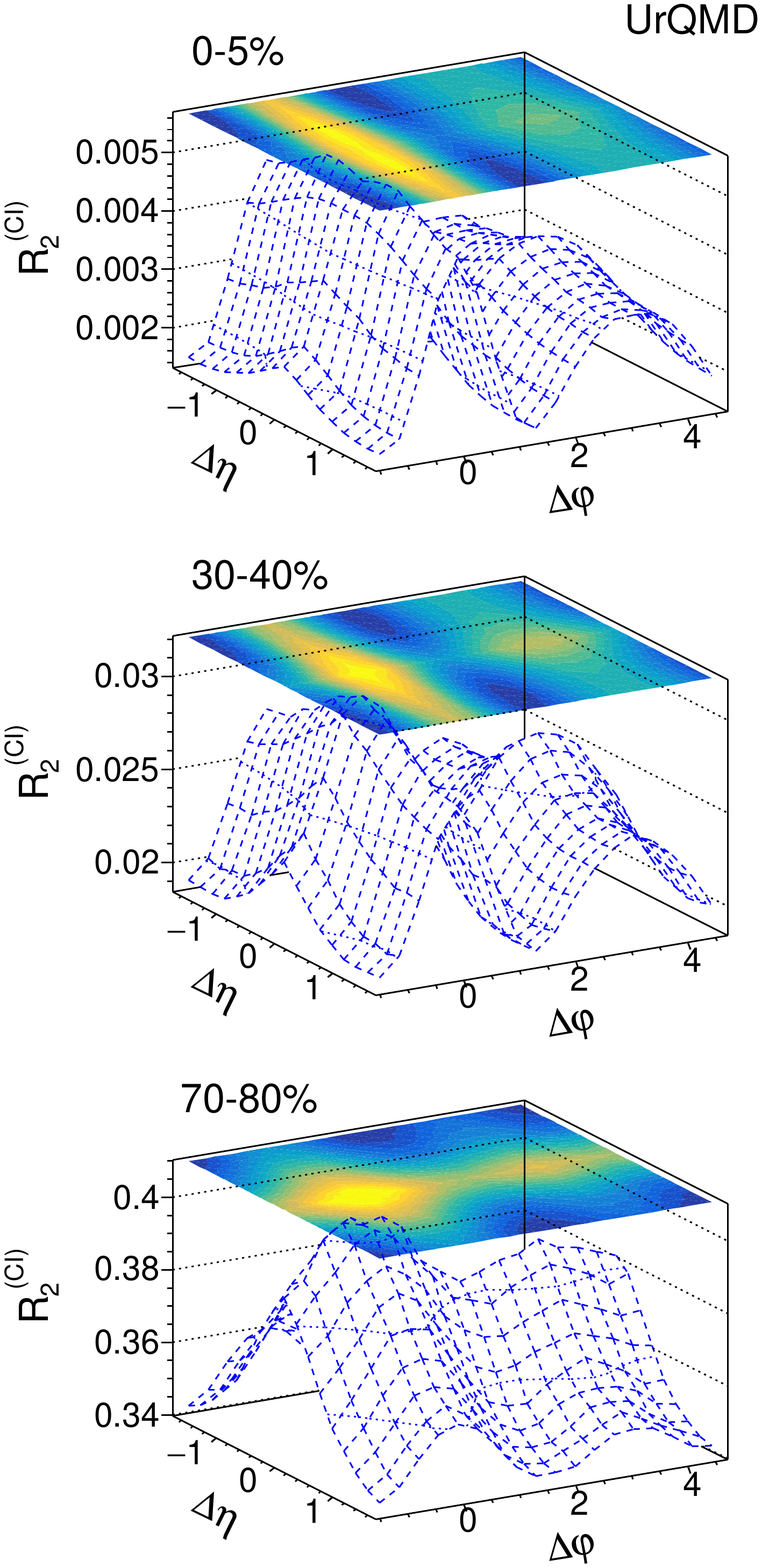}
\includegraphics[scale=0.30,clip=true,trim=101pt 5pt 91pt 1pt]{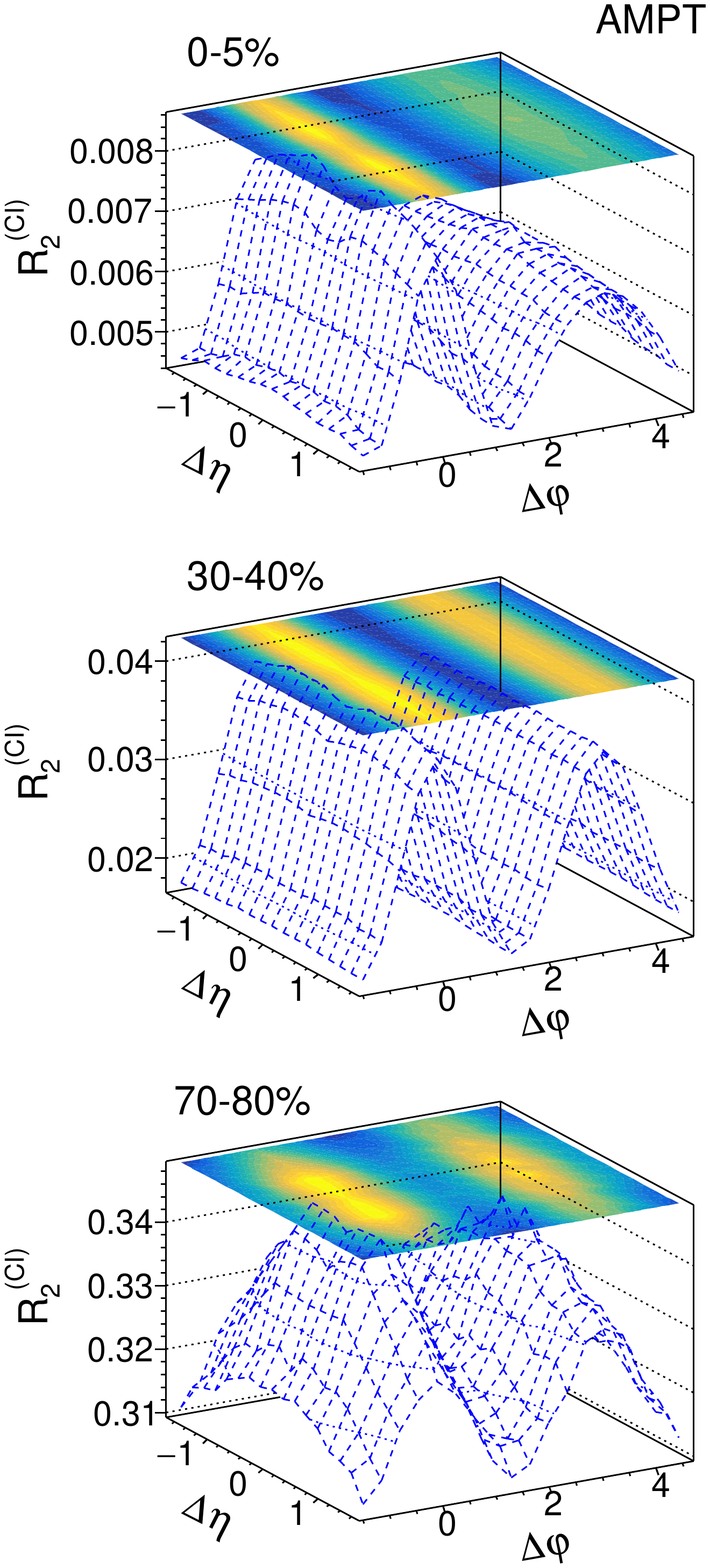}
\includegraphics[scale=0.30,clip=true,trim=101pt 5pt 91pt 1pt]{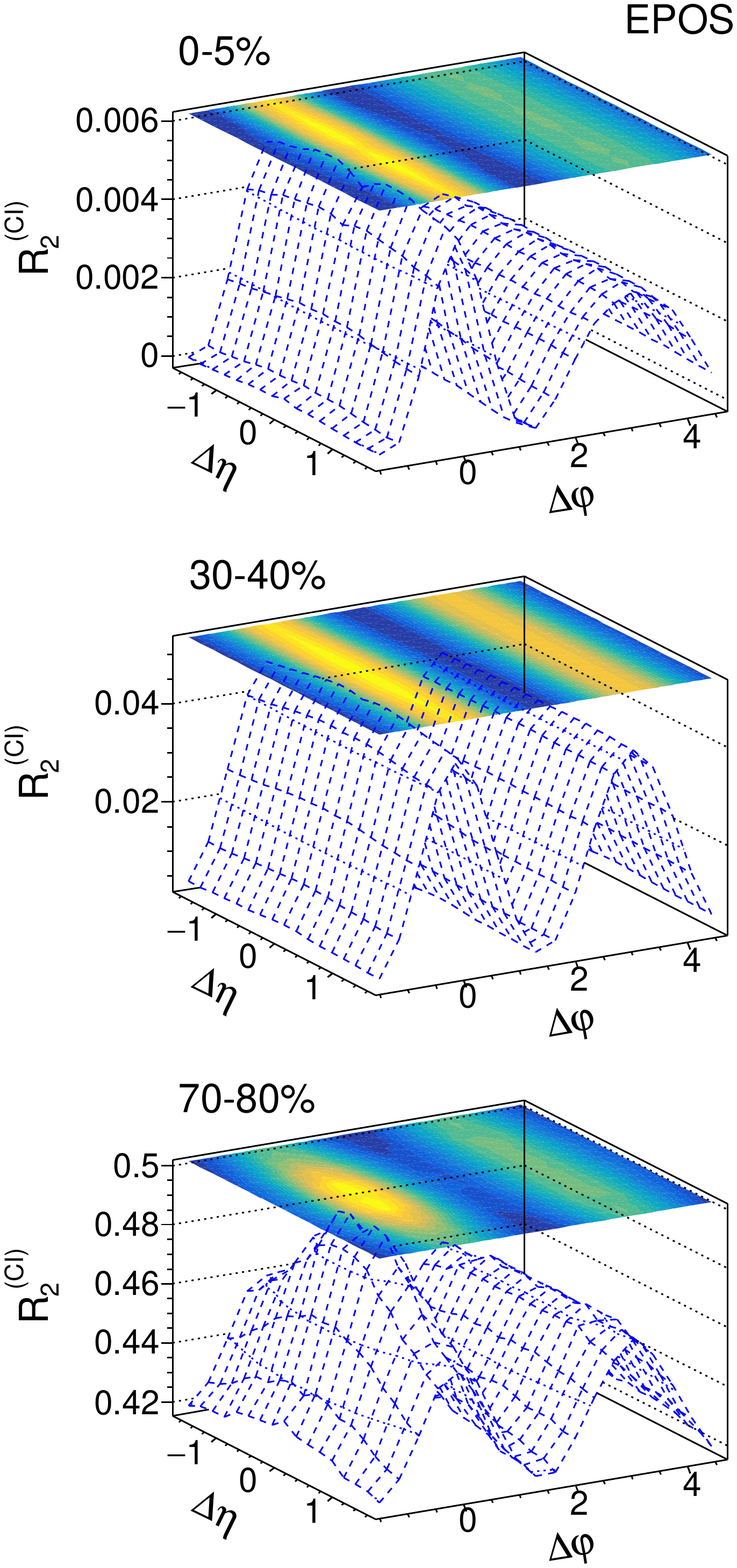}
\includegraphics[scale=0.30,clip=true,trim=75pt 5pt 91pt 1pt]{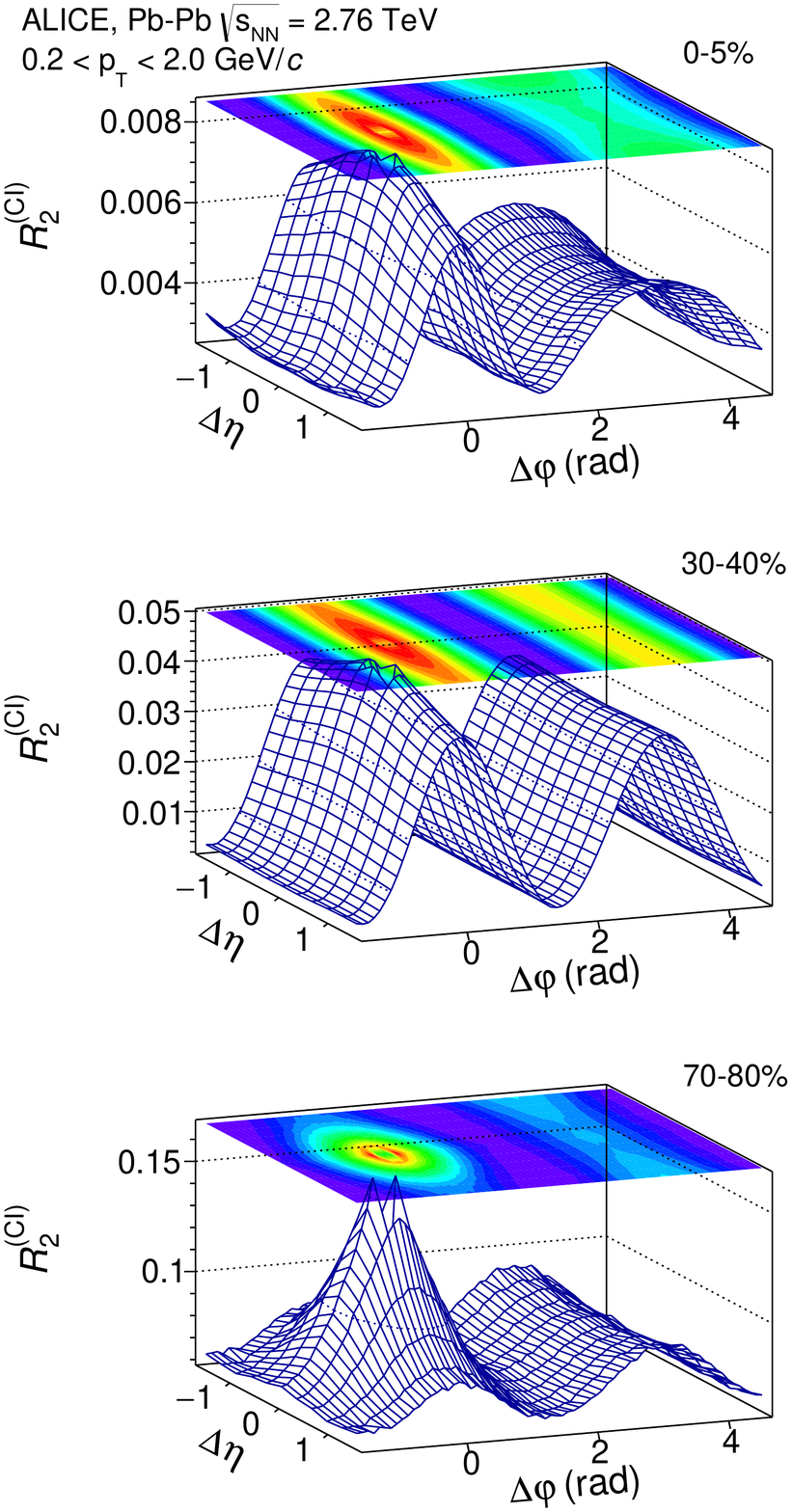}
\caption{Correlators $R_2^{(CI)}$ obtained with the UrQMD, AMPT (SON/RON) and EPOS models compared to correlators measured by the ALICE collaboration~\cite{PhysRevC.100.044903} in \PbPb\  collisions at $\snn = 2.76$  TeV for three representative collision centrality ranges. Correlators are based on charged hadrons in the range $0.2 <  \pt \leq 2.0$~\gevc. See text for details. }
\label{fig:cR2CI}
\end{center} 
\end{figure*}

\begin{figure*}[htb!] 
\begin{center} 
\includegraphics[scale=0.30,clip=true,trim=101pt 5pt 91pt 1pt]{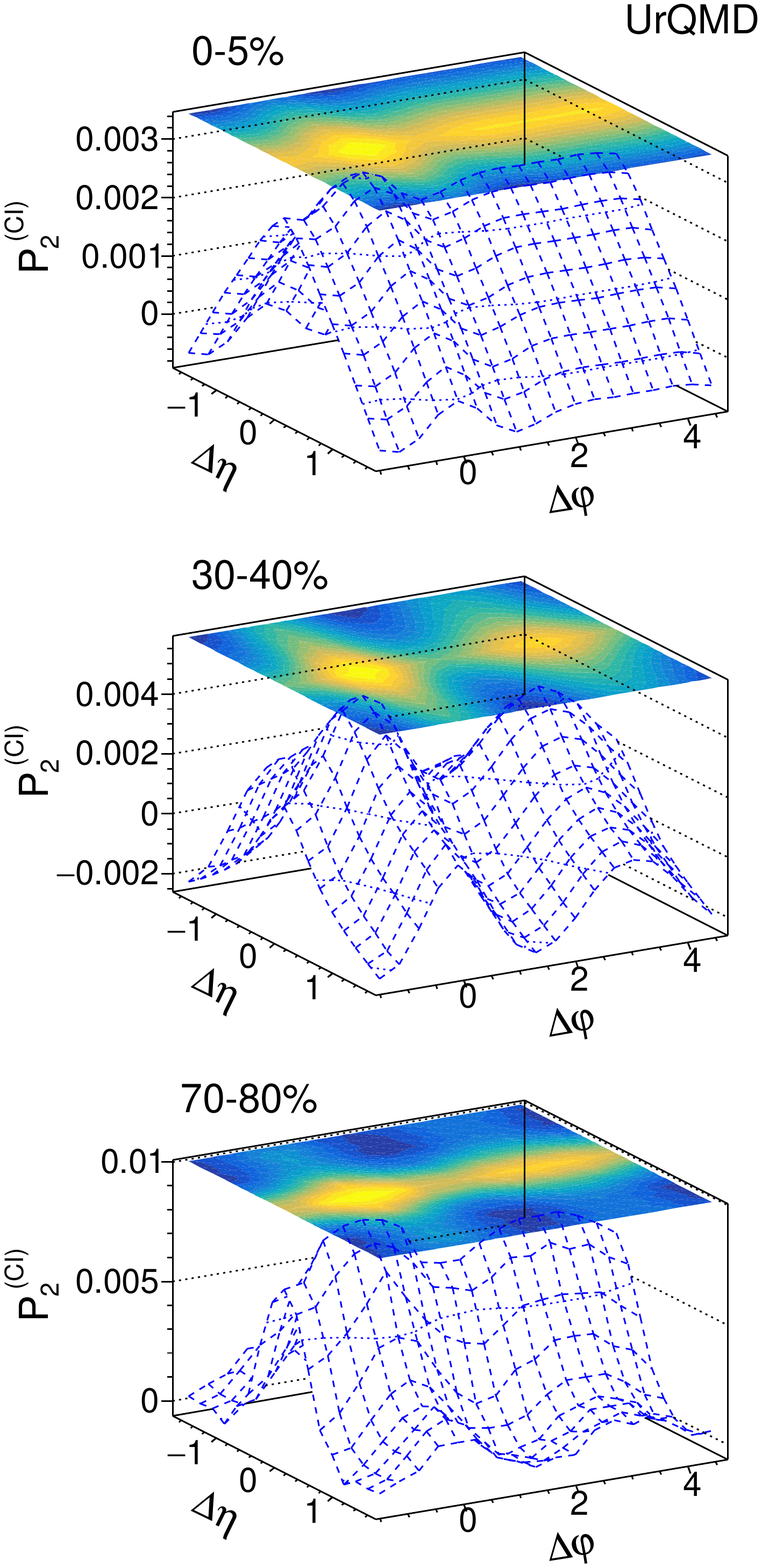}
\includegraphics[scale=0.30,clip=true,trim=101pt 5pt 91pt 1pt]{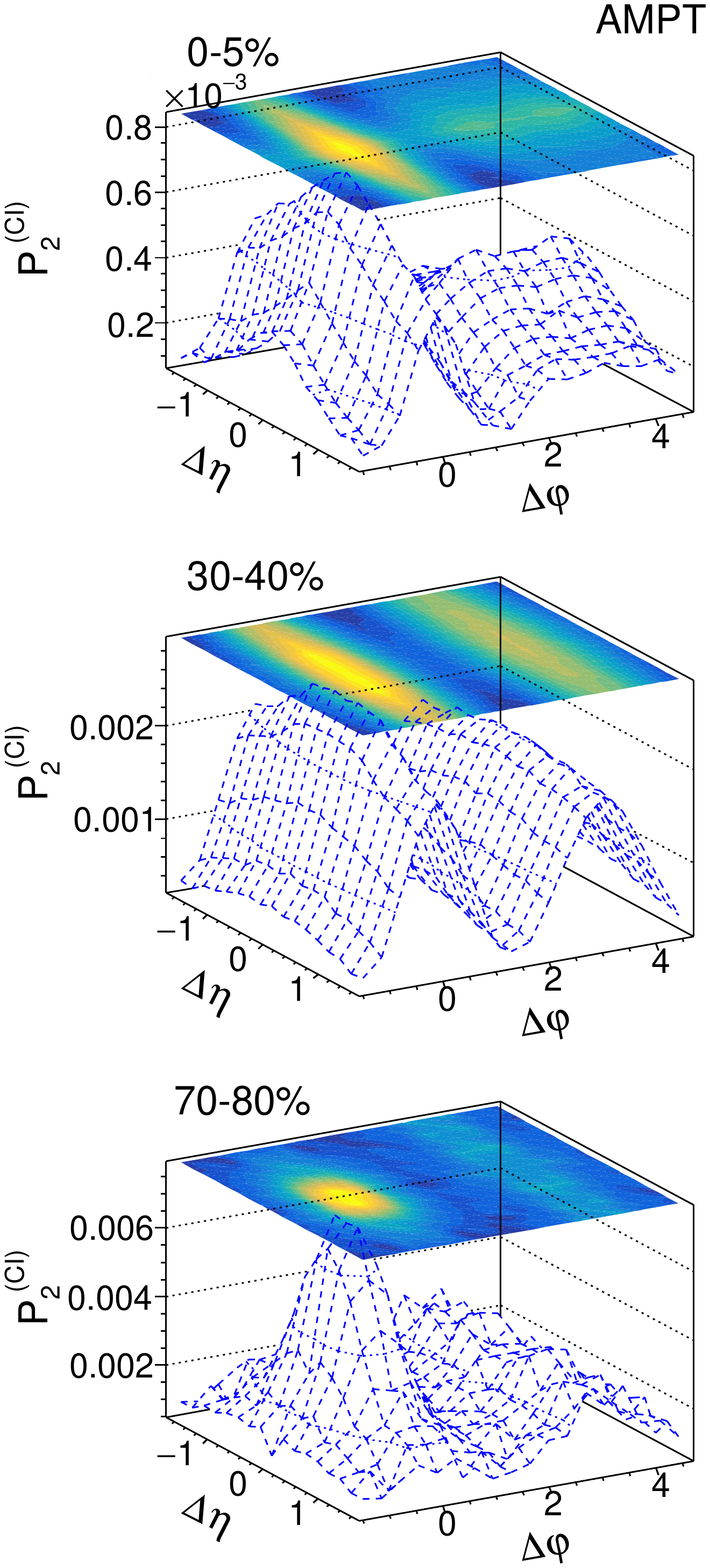}
\includegraphics[scale=0.30,clip=true,trim=101pt 5pt 91pt 1pt]{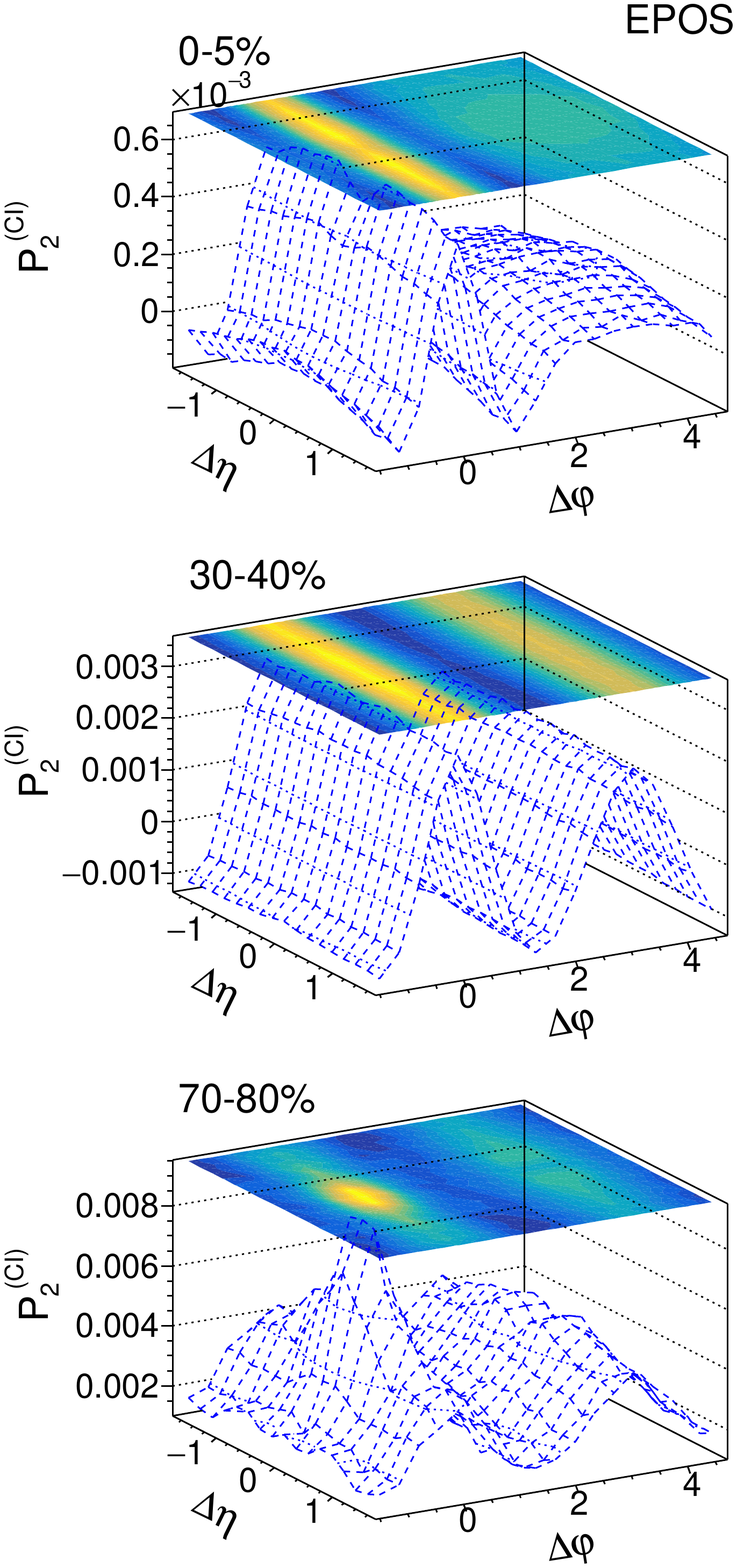}
\includegraphics[scale=0.30,clip=true,trim=75pt 5pt 91pt 1pt]{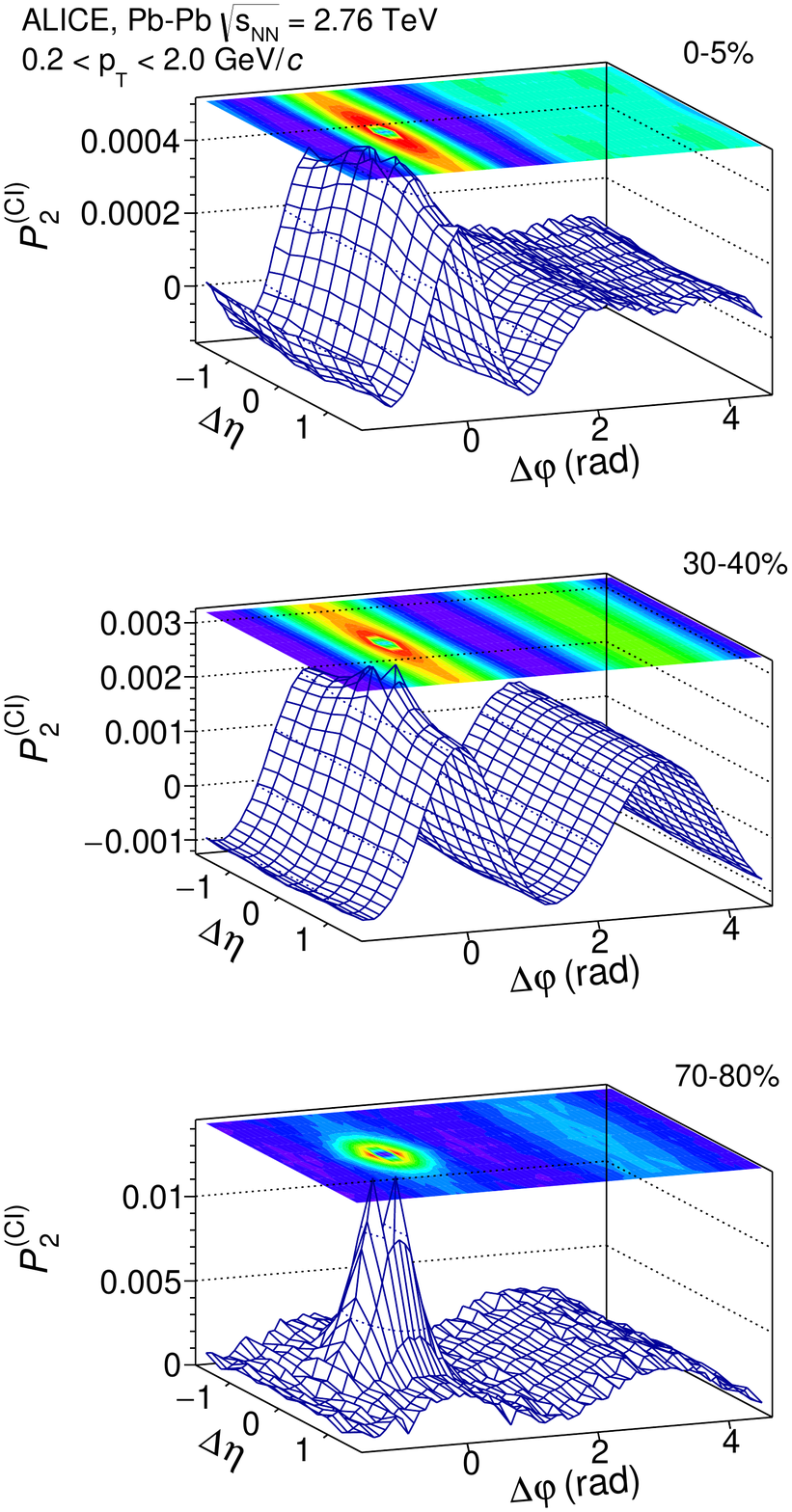}
\caption{Correlators $P_2^{(CI)}$ obtained with the UrQMD, AMPT (SON/RON) and EPOS models compared to correlators measured by the ALICE collaboration~\cite{PhysRevC.100.044903} in \PbPb\  collisions at $\snn = 2.76$  TeV for three representative collision centrality ranges. Correlators are based on charged hadrons in the range $0.2 <  \pt \leq 2.0$~\gevc. See text for details.}
\label{fig:cP2CI}
\end{center} 
\end{figure*}

\begin{figure*}[htb] 
\includegraphics[scale=0.8]{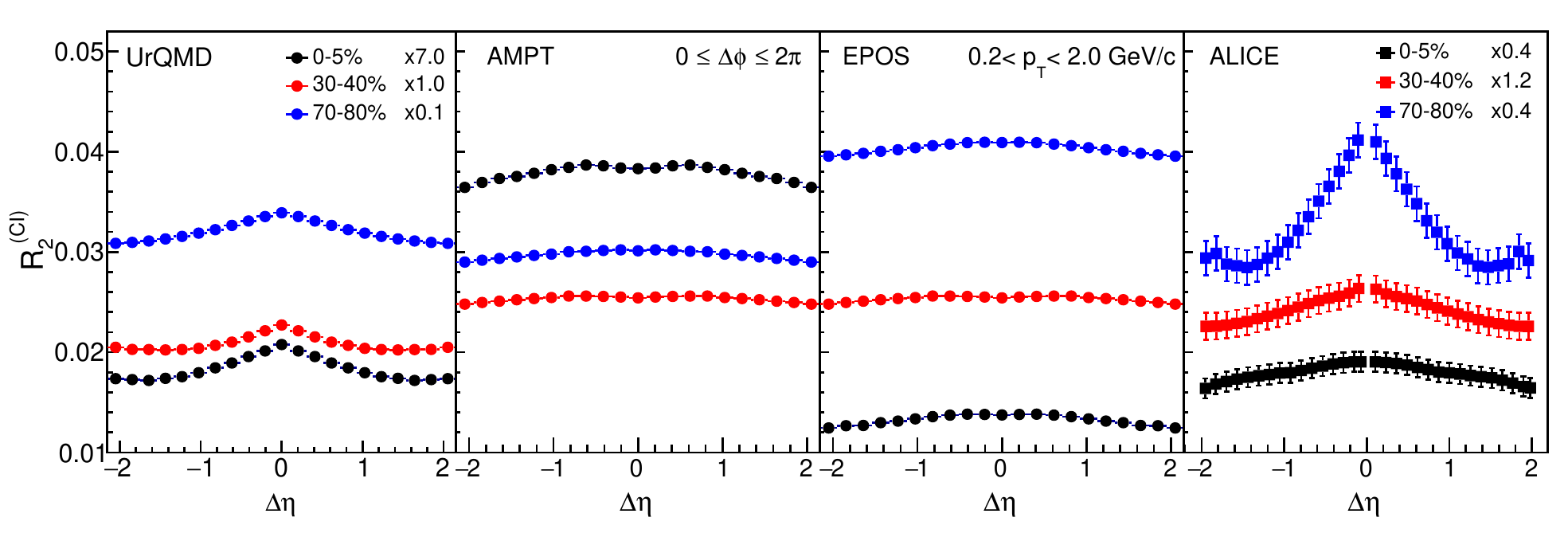}
\includegraphics[scale=0.8]{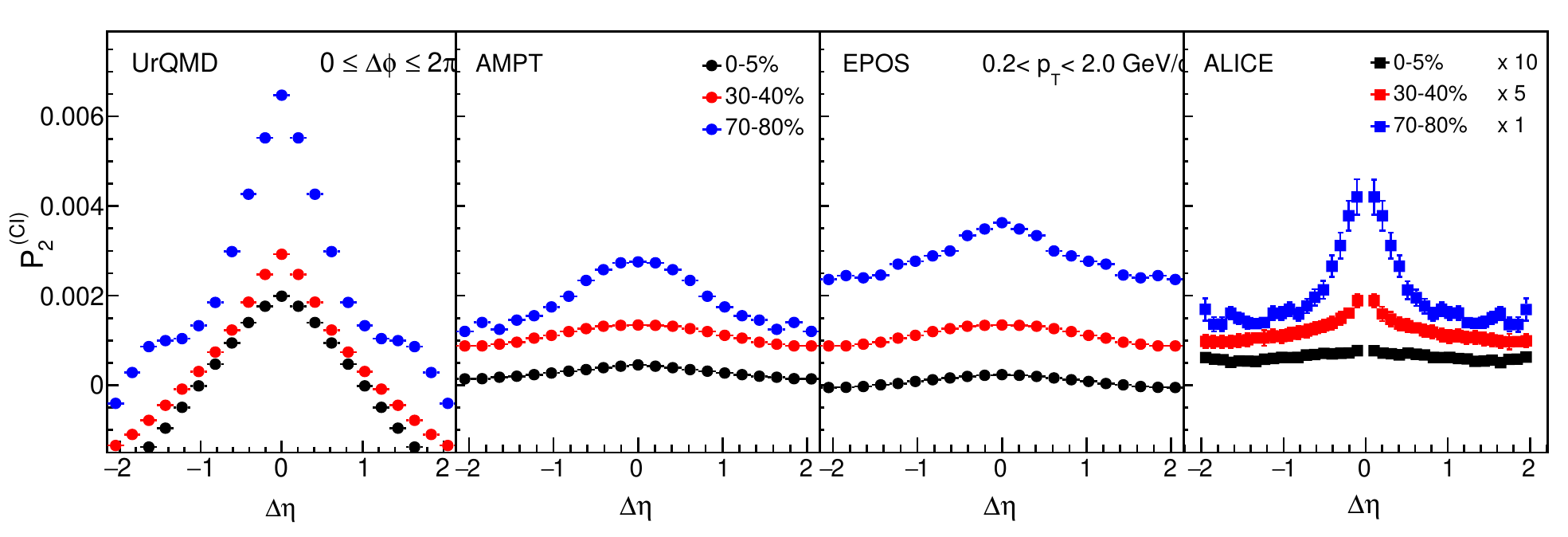}
\caption{Projections of $R_2^{(CI)}$ and $P_2^{(CI)}$ correlators of charged hadrons obtained with UrQMD, AMPT and EPOS event generators compared to projections of the correlators measured by the ALICE collaboration~\cite{PhysRevC.100.044903} in \PbPb\ collisions at $\snn$ = 2.76 TeV  shown in Figs.~\ref{fig:cR2CI} and~\ref{fig:cP2CI}. Scaling factors listed for $R_2$ in the left panel apply to the three model calculations.}
\label{fig:cR2CI_deta}
\end{figure*}

\begin{figure*}[htb!] 
\includegraphics[scale=0.7]{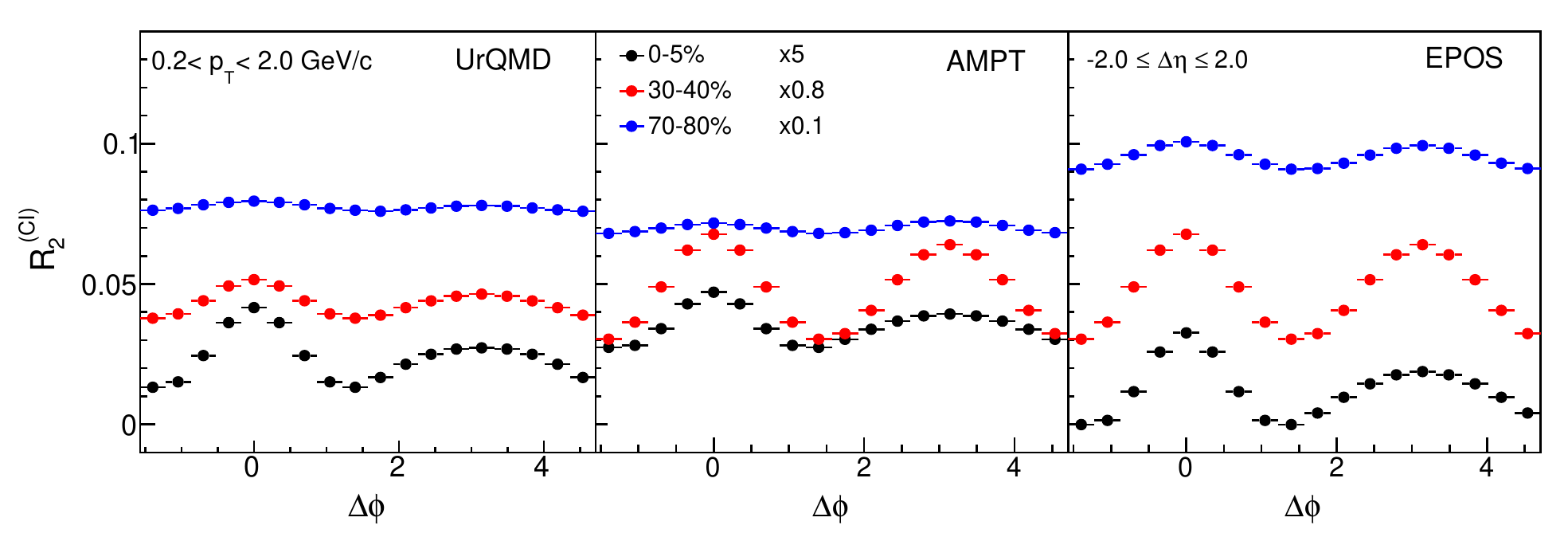}
\includegraphics[scale=0.7]{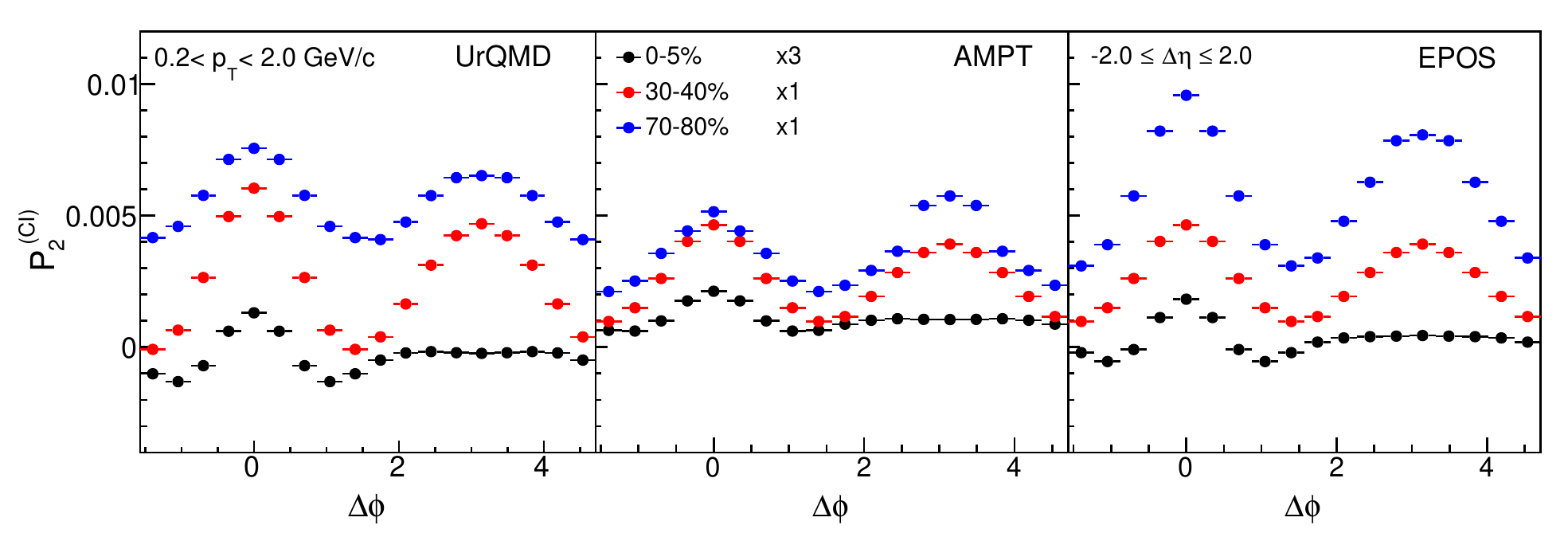}
\caption{Projections of  $\rm{R_2^{(CI)}}$(top) and  $\rm{P_2^{(CI)}}$ (bottom) correlators of charged hadrons in the range  $0.2 < \pt \leq 2.0$ obtained with UrQMD, AMPT and EPOS for \PbPb\ collisions at  $\snn$ = 2.76 TeV. The $\Dphi$  projections are calculated as averages of the two-dimensional correlations in the ranges $|\Deta|<2$. Scaling factors listed  for $R_2$  and $P_2$ in the central panel apply to the three model calculations.}
\label{fig:cR2CI_dphi}
\end{figure*}

 The $R_2$ and $P_2$ correlators obtained in simulations of \PbPb\  collisions at $\snn = 2.76$ TeV, with UrQMD, AMPT, and  EPOS, are compared to ALICE measurements~\cite{Adam:2017ucq,PhysRevC.100.044903} in Figs.~\ref{fig:cR2LS}--\ref{fig:cP2CD} for three representative  multiplicity classes corresponding to 0--5\% (most central collisions), 30--40\% (mid-central collisions) and 70--80\% (peripheral collisions) fractions of the interaction cross section.
For the sake of simplicity, and without sizable bias, the model events were classified based on their impact parameter $b$ following the technique used in Ref.~\cite{PhysRevC.88.044909}. Unfortunately, it was not possible, with the resources available to these authors, to generate model datasets of size comparable to those acquired experimentally by the ALICE collaboration. Some of the simulated correlators presented in this section, particularly the CD correlators, thus suffer 
from limited statistics that somewhat hinder comparisons with experimental data. Our discussion thus mainly
focuses on model calculations for  $R_2$ and $P_2$ CI correlation functions and   $R_2$ CD correlation functions.
 
The model calculations were carried out  with  event and track selection criteria designed  to mimic the data collected by the ALICE collaboration. The analysis was performed on minimum bias events.  Unidentified charged hadrons were selected in the pseudorapidities  range  $|\eta| < 1.0$, the azimuth angle range $0 \le \varphi < 2\pi$,  and transverse momenta  range  $0.2 \le p_{\rm T} \leq 2.0$~GeV/$c$. No other experimental filter were used in the calculation of the correlators given the published ALICE data were already corrected for particle losses (single particle detection efficiency) and given resolution smearing and  contamination from background processes were assessed to be essentially negligible by the ALICE collaboration in their measurements of the  $R_2$ and $P_2$ correlators.

We begin with a discussion of unidentified like-sign (LS) and unlike-sign (US) charged hadron   correlators  in sec.~\ref{sec:ResultsLSUS}. Charge independent (CI) and charge dependent (CD) correlation functions are presented in sec.~\ref{sec:ResultsCI} and~\ref{sec:ResultsCD}, respectively.  We shall examine, in particular, whether the $R_2$ and $P_2$ correlators obtained with the three models feature the azimuthal modulations, near-side peak, and away-side ridge structures observed in measured correlation functions reported by the ALICE collaboration~\cite{Adam:2017ucq,PhysRevC.100.044903}. 

\subsection{LS, US  correlation functions} 
\label{sec:ResultsLSUS}

LS and US $R_2$  correlators obtained with  UrQMD, AMPT, and  EPOS  are compared to ALICE measurements in Figs.~\ref{fig:cR2LS} and \ref{fig:cR2US}, respectively.

The measured  LS and US $R_{2}(\Delta\eta,\Delta\varphi)$ exhibit  similar features and evolution with collision centrality. Both correlators feature a somewhat narrow near-side peak, i.e., a peak centered at $(\Delta\eta,\Delta\varphi)=(0,0)$, in peripheral collisions (70-80\%). The amplitude of this peak decreases while a strong $\Delta\varphi$ modulation, associated with anisotropic flow, emerges in more central collisions.  A near-side peak with small amplitude remains in US correlations measured in most central collisions while a small depression replaces it in LS correlations. One also notes that both LS and US correlators feature a bowed dependence on $\Delta\eta$ on the away-side, i.e., for $\Delta\varphi \sim \pi$.

At first glance, it is remarkable to observe that the UrQMD, AMPT, and EPOS models qualitatively reproduce the strength and salient components of the measured correlation functions, particularly in the  0--5\% and 30--40\% centrality ranges. We find, indeed, that  the  models capture several  of the features  seen in the data, including the observed diminishing correlation strength  observed with collision centrality. That alone, in fact, constitutes a great measure of success for the models. Some puzzling differences are however observed, which we proceed to discuss.  For instance, all three models produce a near-side peak in LS and US correlators but have  varying successes in reproducing the centrality evolution of its amplitude and shape in more central collisions. In particular, the UrQMD model, additionally, yields an extraneous $\Delta\varphi$ ridge at $\Delta\eta=0$ in the three centrality ranges considered. The three models qualitatively reproduce the presence of $\Delta\varphi$ modulations and feature some collision centrality dependence but they do not strictly match the trend observed in the data. 

They  also produce correlation strengths that are a factor of 3 to 5 too large in peripheral collisions.   Additionally, one observes  that the three models qualitatively reproduce the presence of the dip at $(\Delta\eta,\Delta\varphi)=(0,0)$  in central collisions in LS correlations but also introduce it in US correlations. Interestingly, AMPT and EPOS yield  such a dip at all centralities for LS pairs.   The  weak  strength of the near-side peak, relative to the away-side correlation amplitude, seen in LS and US correlations measured in peripheral collisions, is an indicator that neither of these models  entirely capture the detailed dynamics of particle production in A-A collisions.

\subsection{Charge Independent (CI) Correlation Functions} 
\label{sec:ResultsCI}

The CI correlators constitute inclusive signatures of the particle
production dynamics and the evolution of the collision system formed
in \PbPb\ interactions. As averages of the US and
LS distributions,  they combine many of the characteristics of  these correlation functions. 
Calculations of the $R_2$ and $P_2$ CI correlators with the UrQMD, AMPT, and  EPOS models for  \PbPb\  collisions at $\snn = 2.76$ TeV are compared to ALICE measurements~\cite{Adam:2017ucq,PhysRevC.100.044903} in Figs.~\ref{fig:cR2CI}--\ref{fig:cP2CI}.  Selected projections of these correlators onto $\Delta\eta$ are shown in  Fig.~\ref{fig:cR2CI_deta}, while projections onto $\Delta\varphi$ are displayed in Fig.~\ref{fig:cR2CI_dphi}. 

As for the more detailed US and LS correlators, one finds that the model calculations capture the  decrease in correlation magnitude of $R_2^{\rm (CI)}$  observed experimentally for increasing event multiplicity (from 70-80\% to 0-5\% collision centrality). As already pointed out above, UrQMD adds an unobserved ridge-like structure vs. $\Delta\varphi$ at $\Delta\eta=0.0$ that contributes considerably to the differences with respect to the data.
This ridge-like structure may be related to the shape of the charged particle pseudorapidity density close to midrapidity observed with UrQMD in its hybrid mode within the transverse momentum range used in this analysis.
One also finds that AMPT and EPOS qualitatively reproduce the emergence of strong $\Delta\varphi$ modulations in mid- to central-collisions but
neither of these models reproduce the correct correlation strength, the bowed dependence on $\Delta\eta$ at $\Delta\varphi=\pi$, or the shape of the near-side peak in most-peripheral collisions. Additionally note that the models produce a relative away-side strength that exceeds that observed in the data. Finally, and as seen in Fig.~\ref{fig:cR2CI_deta}, the three models  do not properly reproduce the  pseudorapidity dependence of the measured $R_2^{\rm (CI)}$ correlators.

Comparison of the model calculations for $P_2^{\rm (CI)}$ are also rather interesting. One finds that EPOS qualitatively reproduces the  narrowness of the near-side peak of $P_2^{\rm (CI)}$ relative to that observed in $R_2^{\rm (CI)}$, as well as the  strong $\Delta\varphi$ modulations measured in 30-40\% and 0-5\%. It also qualitatively replicates the observed dip measured at $(\Delta\eta,\Delta\varphi)=(0,0)$ in most central collisions.  The computed shape of the away-side is however somewhat incompatible with that observed in the data, possibly owing  to a mismatch of the harmonic flow coefficient dependence on $\Delta\eta$. We study this question in more detail later in this section.
Switching the focus to AMPT's calculations, one finds that this model also qualitatively reproduces the relative narrowness of the near-side of $P_2^{\rm (CI)}$ compared to that of $R_2^{\rm (CI)}$. It also qualitatively reproduces the presence of strong $\Delta\varphi$ harmonics. However, AMPT yields a very steep dependence on $\Delta\eta$ on the away-side of  $P_2^{\rm (CI)}$, in most central collisions,  which is in clear disagreement with the measured data.   Note that the UrQMD model  produces  such a steep dependence on $\Delta\eta$  at all collision centralities which relativizes the strong $\Delta\varphi$ modulations observed in mid- to central-collisions. Moreover, UrQMD  shows similar amplitudes on the away and near side  at all centralities, that are not observed experimentally.

\begin{figure*}[htb!] 
\includegraphics[scale=0.8,clip=true,trim=100pt 200pt 100pt 200pt]{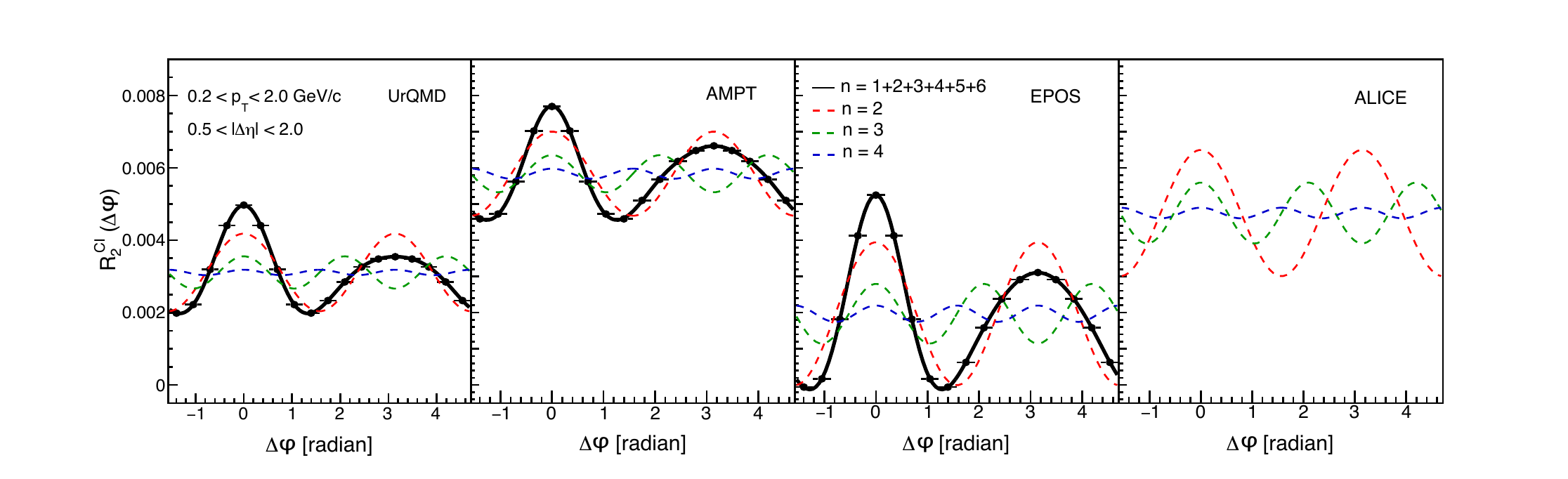}
\includegraphics[scale=0.8,clip=true,trim=100pt 200pt 100pt 200pt]{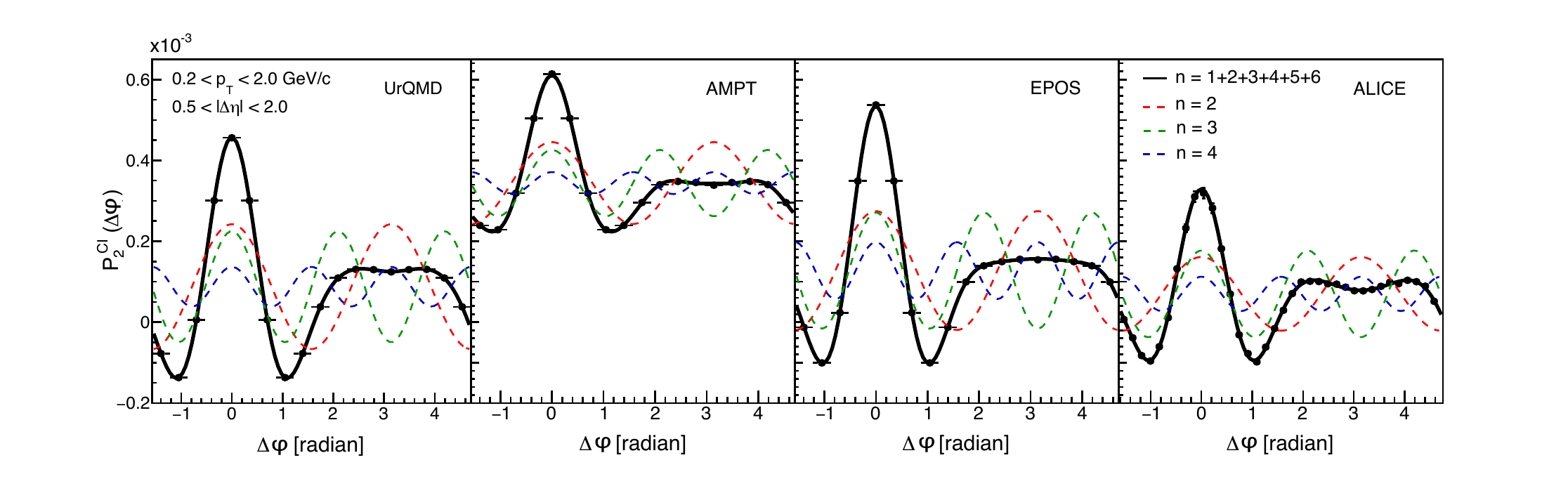}
\caption{Fourier decompositions of projections of the  $\rm{R_2^{(CI)}}$ (top) and $\rm{P_2^{(CI)}}$ (bottom) correlators of charged hadrons in the range  $0.2 < \pt \leq 2.0$ obtained in 5\% most central collisions simulated with UrQMD, AMPT and EPOS and 5\% most central collisions measured  by the ALICE collaboration. Solid lines: Fourier decomposition  fits calculated to the 6th order; dash lines: $n=$2, 3, and 4 components obtained in the fits.  The ALICE collaboration did not report $\Delta\varphi$ projections for $\rm{R_2^{(CI)}}$~\cite{PhysRevC.100.044903}. Plotted is the $\Delta\varphi$ dependence of the $n=$2, 3, and 4 Fourier components estimated  from published values of the flow coefficients $v_2$, $v_3$, and $v_4$~\cite{Adam:2017ucq}.}
\label{fig:cDphiCI}
\end{figure*}  

\begin{figure*}[htb]  
\includegraphics[scale=0.41,clip=true,trim=1pt 0pt 5pt 0pt5]{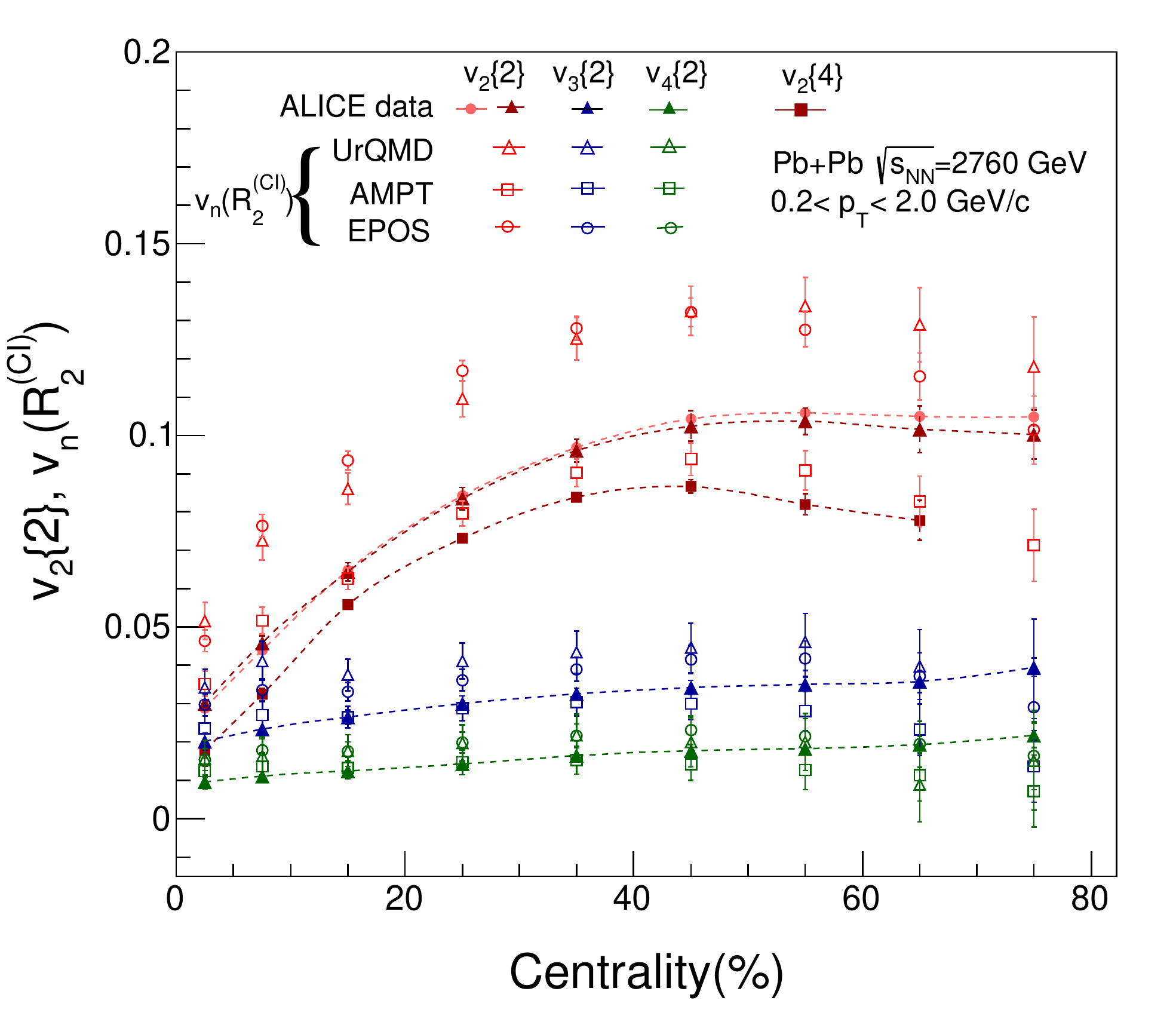}
\includegraphics[scale=0.4,clip=true,trim=1pt 0pt 5pt 0pt4]{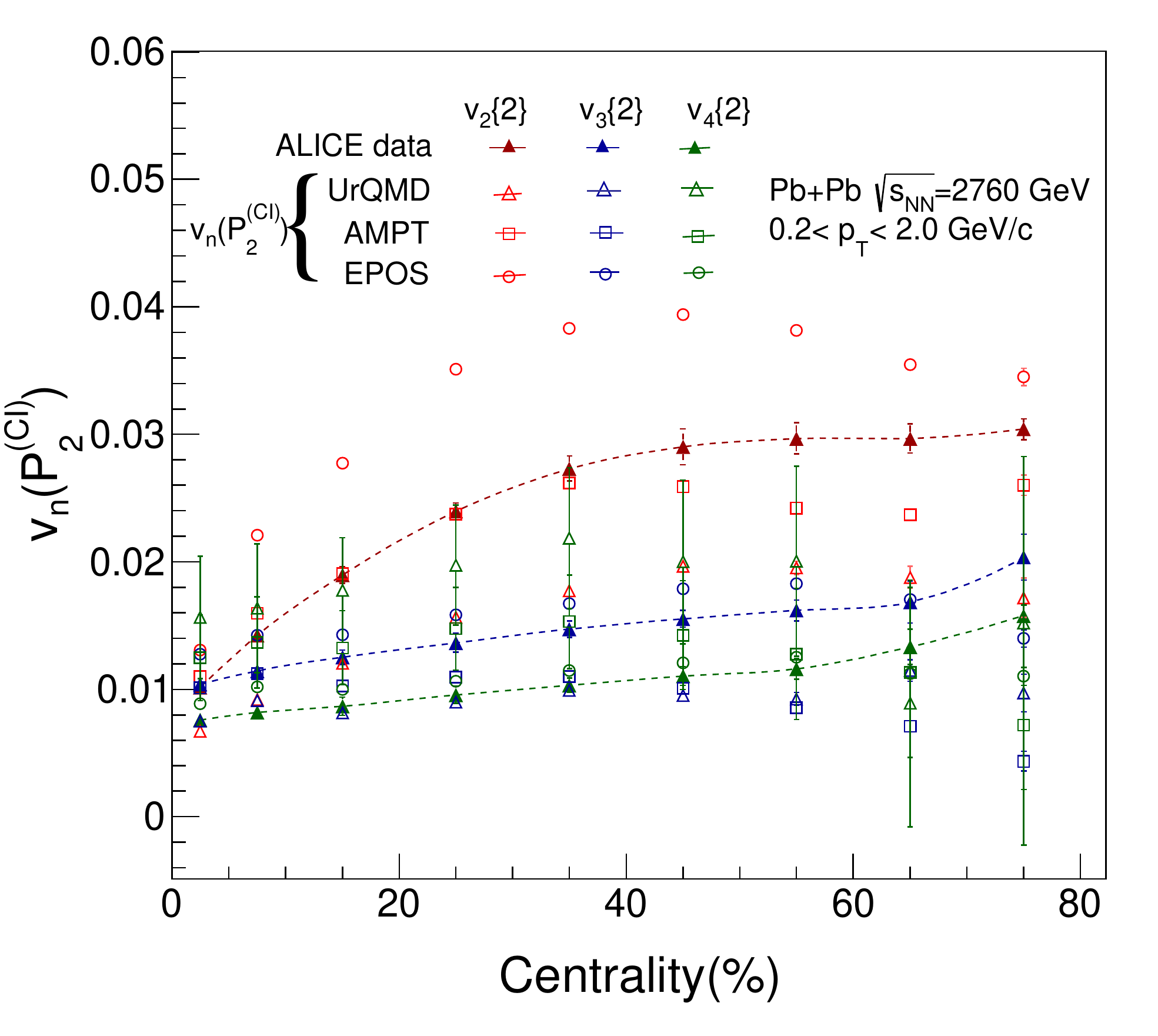}
\caption{Fourier coefficients $v_{n}$, with n=2,3,4,
extracted from  $\rm{R_2^{(CI)}}$ (left) and $\rm{P_2^{(CI)}}$ (right) correlation functions.}
\label{fig:cR2CI_Fourier}
\end{figure*}  

Let us  further examine the model calculations for the $R_2^{\rm (CI)}$ and $P_2^{\rm (CI)}$ correlators shown and compared to ALICE data in Fig.~\ref{fig:cR2CI} and~\ref{fig:cP2CI}, respectively. Both the measured $R_2^{\rm (CI)}$ and $P_2^{\rm (CI)}$ correlation functions  exhibit  a $\Delta\varphi$ modulation  that extends across the full $\Delta\eta$ range of the ALICE TPC acceptance.   We thus focus the discussion on this modulation by plotting projections of the  calculated correlators onto the  $\Dphi$ axis  in Fig.~\ref{fig:cR2CI_dphi}.   First considering the $R_2^{\rm (CI)}$  projections, one finds that the   three models produce average correlation strengths and $\Delta\varphi$ modulations that evolve with collision centrality, but produce  average magnitudes  and   modulation amplitudes   that appear to be  mutually distinct and in quantitative disagreement with the measured data. We elaborate on this point by performing a Fourier decomposition, $F(\Dphi) = a_0 + 2 \sum_{n=1}^{6} a_n~\cos(n\Dphi)$,  of the computed correlation functions. The  functions $F(\Dphi)$ obtained from the fits, and the four lower order components,  are shown for both correlators and the three models in Fig.~\ref{fig:cDphiCI}, along with 
results of similar fits carried out   on published ALICE data~\cite{ALICE:2011ab}. The magnitude of the $v_n = \sqrt{a_n}$ coefficients obtained from  the fits are shown  as a function of collision centrality in Fig.~\ref{fig:cR2CI_Fourier}. We find the AMPT calculations for $v_2(R_2^{\rm (CI)})$ have a magnitude between 
those of $v_2\{2\}$ and $v_2\{4\}$ reported by the ALICE collaboration~\cite{Abelev:2014mda}, in qualitative agreement with the 
magnitude of $v_2$ expected when flow fluctuations and non-flow effects are suppressed.  We find that AMPT also produces $v_3\{2\}$ and $v_4\{4\}$ coefficient magnitudes in very good agreement to values reported by the ALICE collaboration. In contrast,  EPOS  tend to systematically overestimate all of the measured coefficients, whereas UrQMD somewhat overestimates the $v_2$ and $v_3$ coeffients but reproduces the $v_4$ coefficients rather well.  Note however that the magnitude $v_n$ coefficients computed with hydrodynamics models is quite  sensitive to the magnitude of the viscosity 
used in the calculations. The UrQMD calculation presented is based on hydrodynamics. It is thus expected that inclusion  of finite viscosities  in the UrQMD calculations could reduce differences with the observed data.

\begin{figure*}[htb] 
\begin{center} 
\includegraphics[scale=0.30,clip=true,trim=101pt 5pt 91pt 1pt]{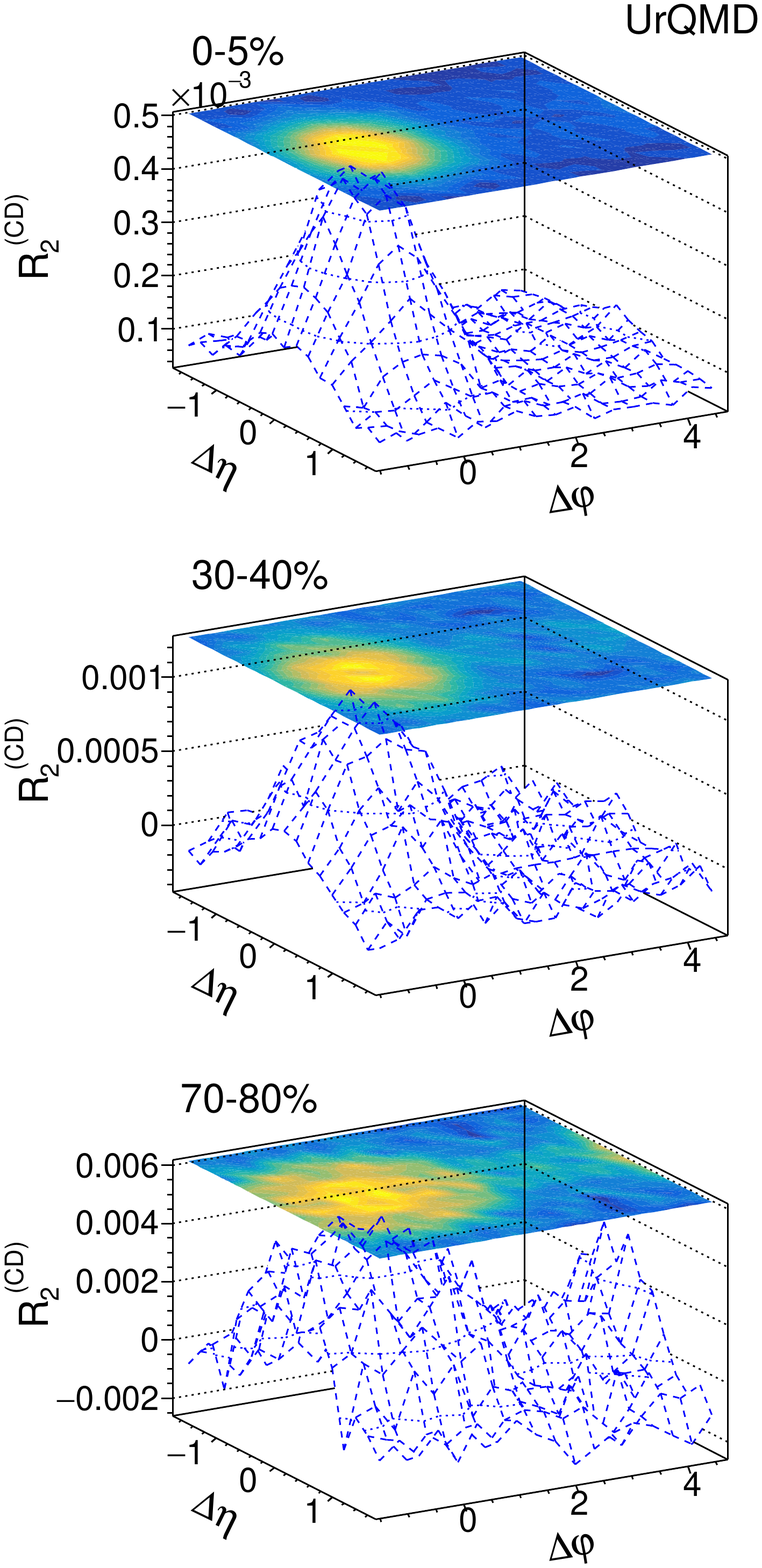}
\includegraphics[scale=0.30,clip=true,trim=101pt 5pt 91pt 1pt]{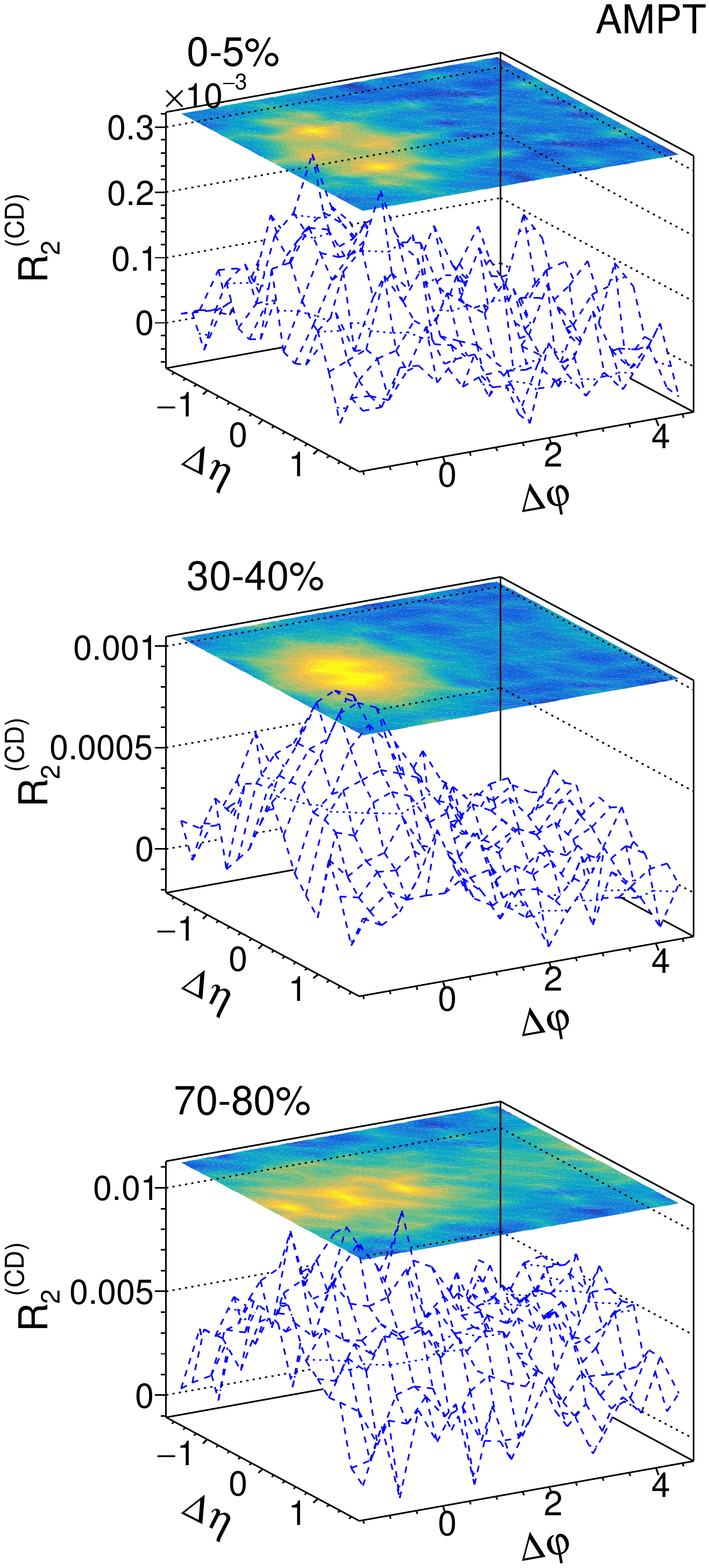}
\includegraphics[scale=0.30,clip=true,trim=101pt 5pt 91pt 1pt]{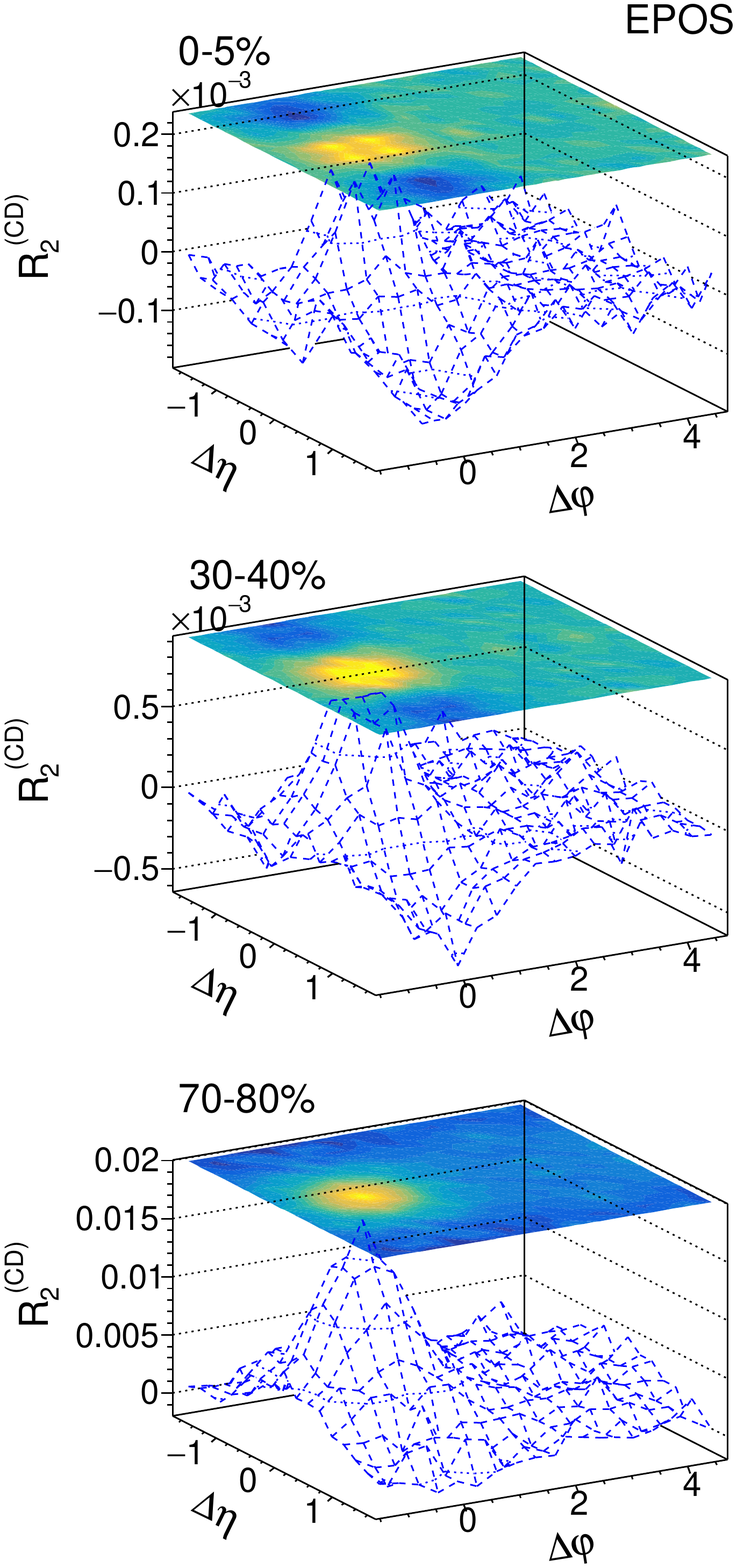}
\includegraphics[scale=0.30,clip=true,trim=75pt 5pt 91pt 1pt]{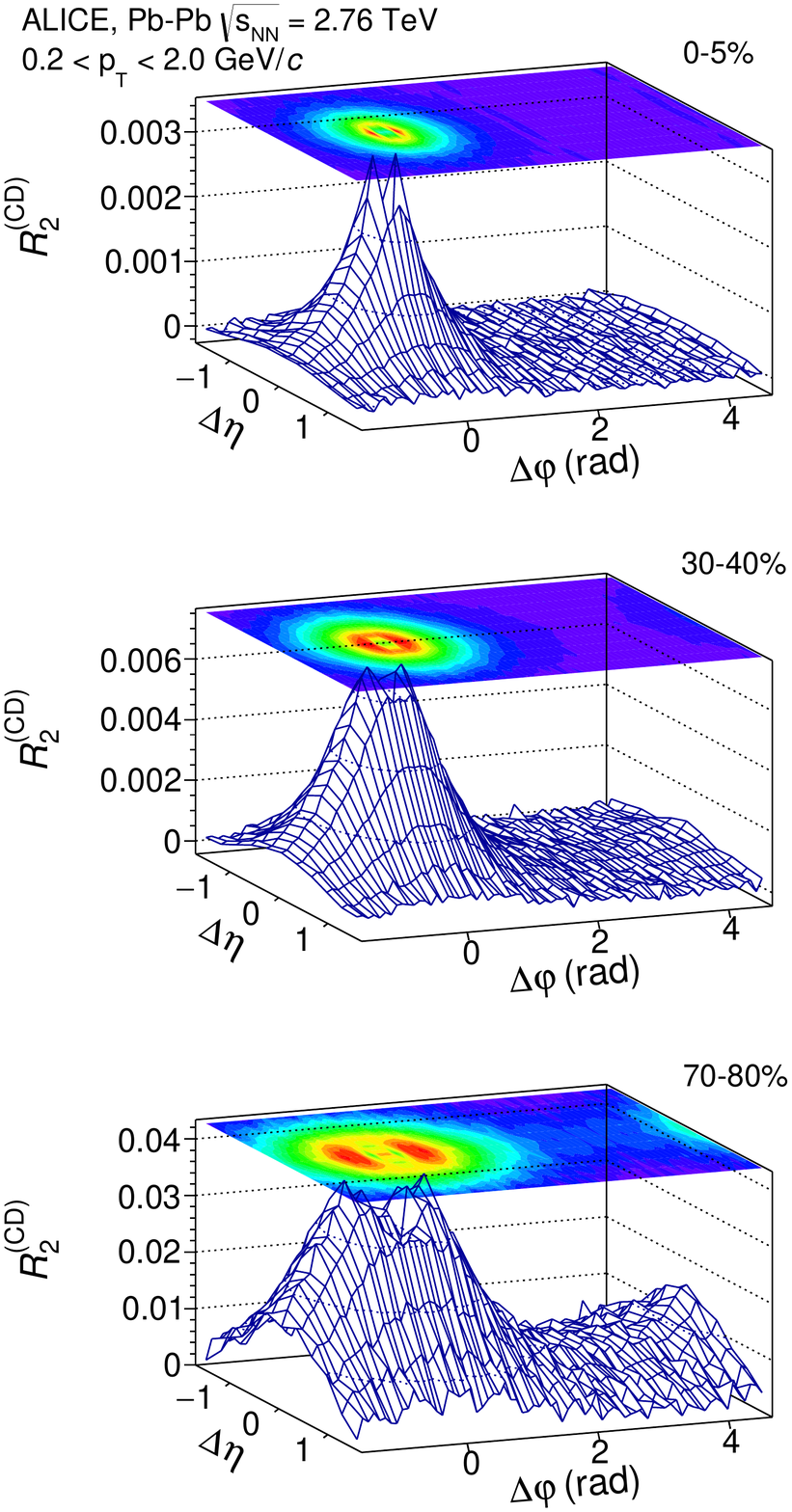}
\caption{Correlators $R_2^{\rm (CD)}$ produced with the UrQMD, AMPT (SON/RON) and EPOS models compared to correlators measured by the ALICE collaboration~\cite{PhysRevC.100.044903} in \PbPb\  collisions at $\snn = 2.76$  TeV for three representative collision centrality ranges. Correlators are based on charged hadrons in the range $0.2 <  \pt \leq 2.0$~\gevc. See text for details.}
\label{fig:cR2CD}
\end{center} 
\end{figure*}

\begin{figure*}[htb!] 
\begin{center} 
\includegraphics[scale=0.30,clip=true,trim=101pt 5pt 91pt 1pt]{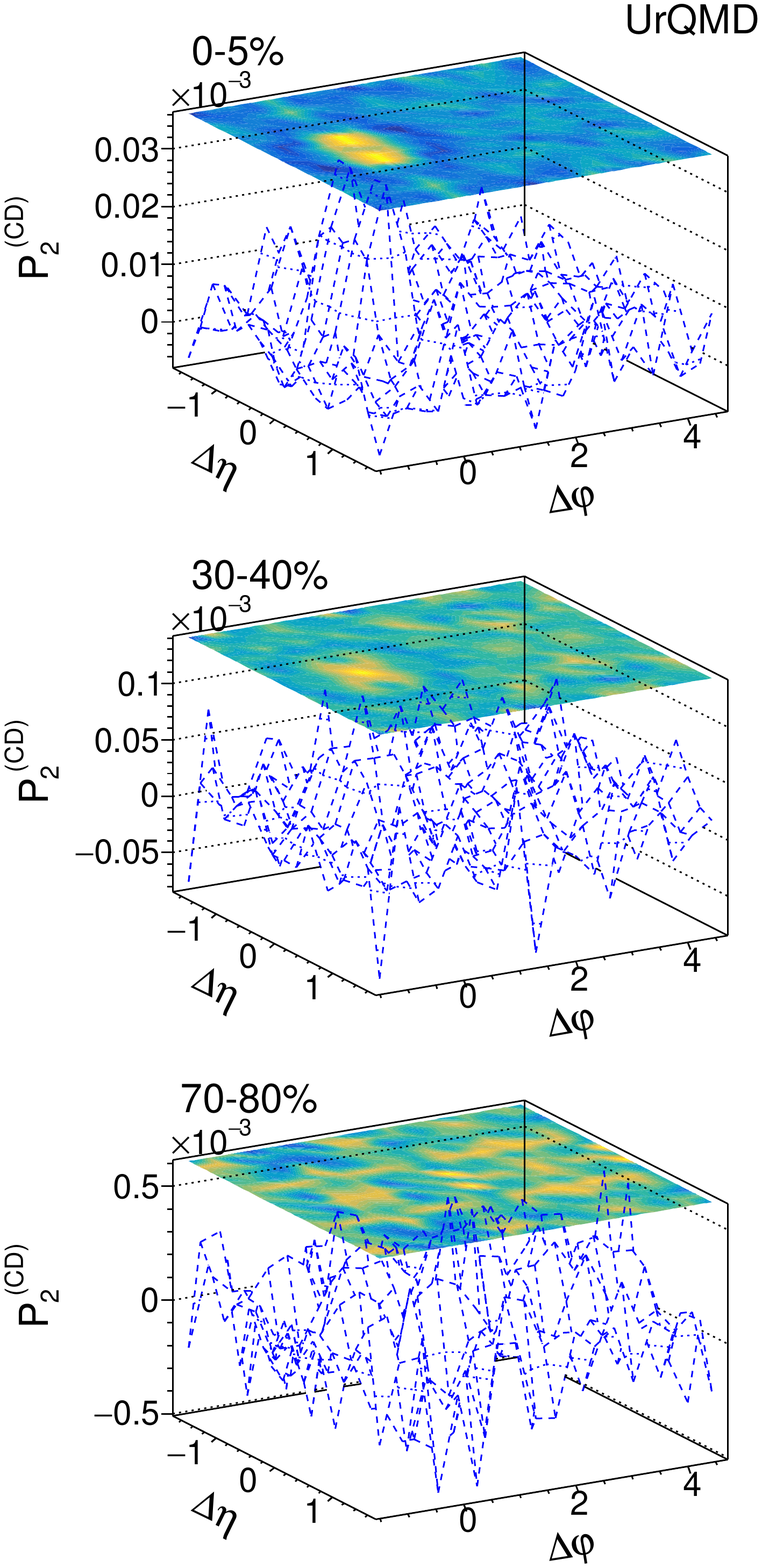}
\includegraphics[scale=0.30,clip=true,trim=101pt 5pt 91pt 1pt]{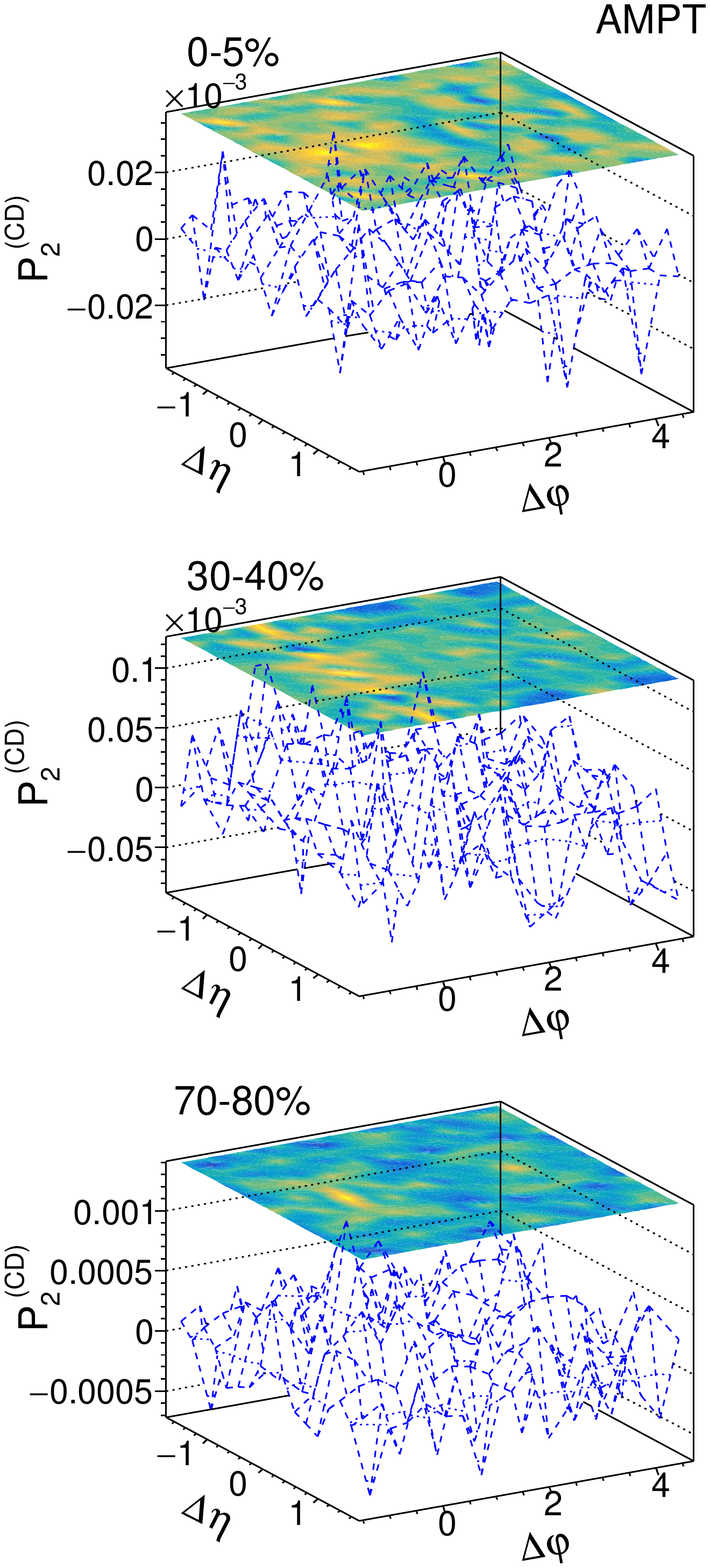}
\includegraphics[scale=0.30,clip=true,trim=101pt 5pt 91pt 1pt]{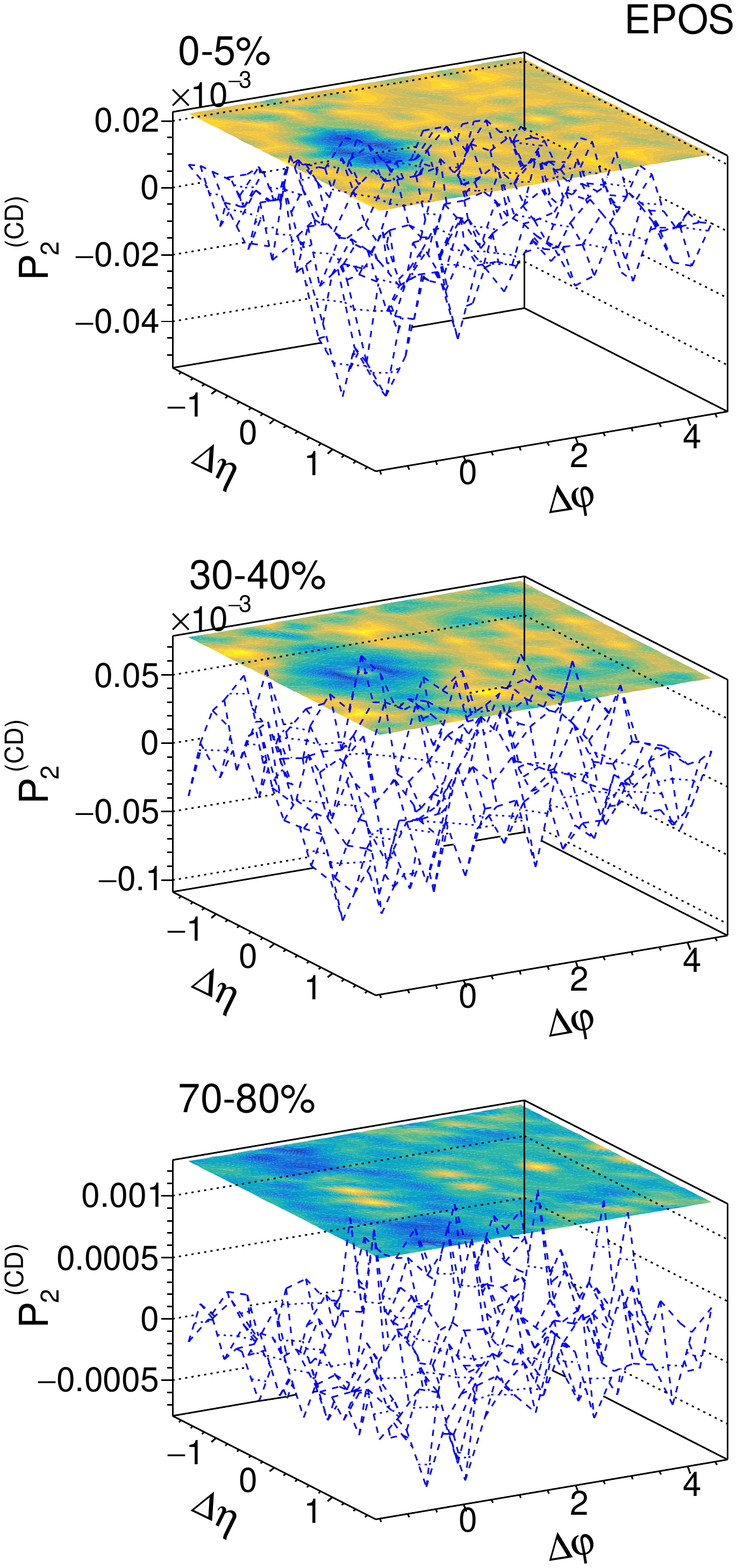}
\includegraphics[scale=0.30,clip=true,trim=75pt 5pt 91pt 1pt]{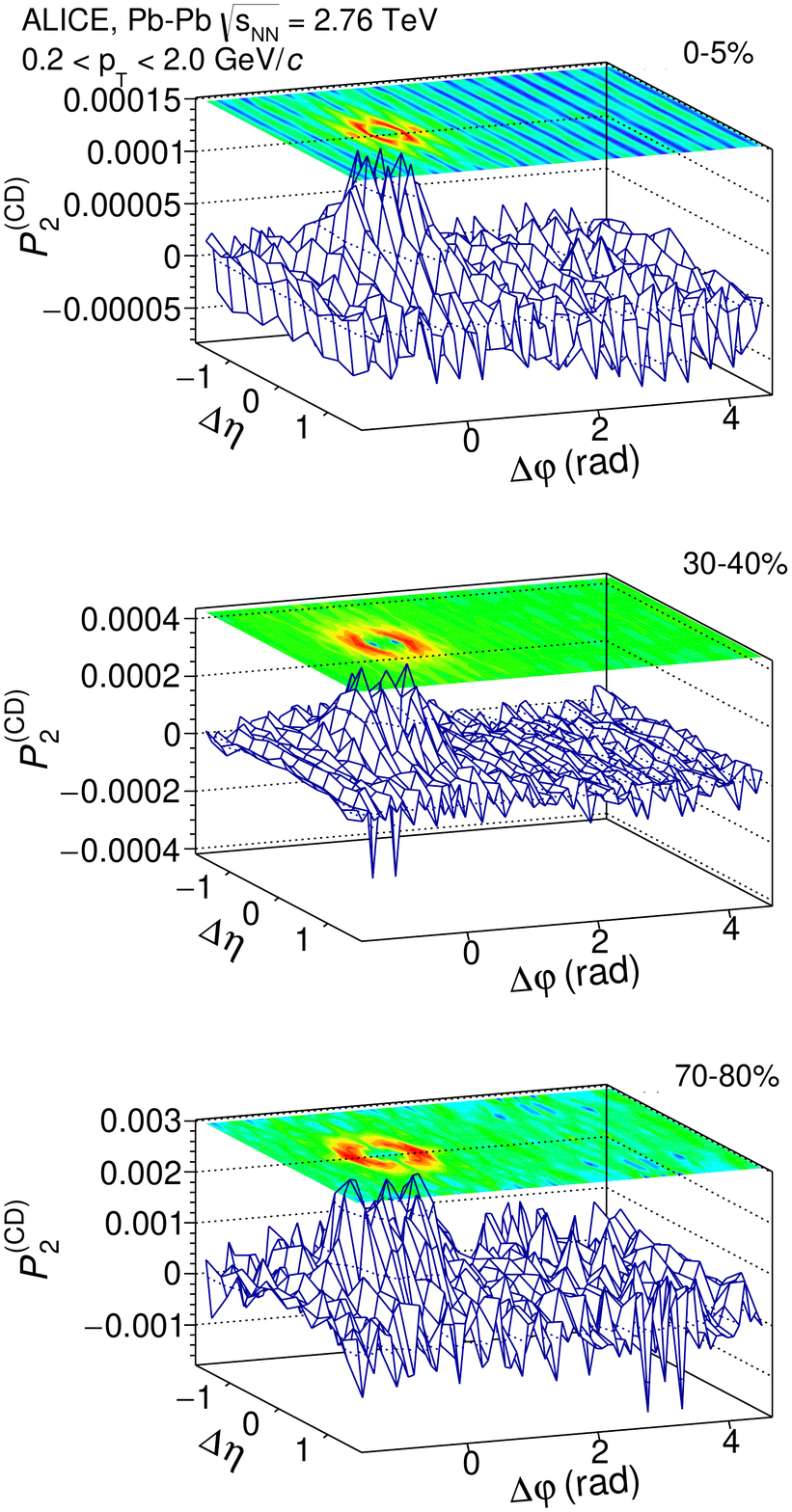}
\caption{Correlators $P_2^{\rm (CD)}$ produced with the UrQMD, AMPT (SON/RON) and EPOS models compared to correlators measured by the ALICE collaboration~\cite{PhysRevC.100.044903} in \PbPb\  collisions at $\snn = 2.76$  TeV for three representative collision centrality ranges. Correlators are based on charged hadrons in the range $0.2 <  \pt \leq 2.0$~\gevc. See text for details.}
\label{fig:cP2CD}
\end{center} 
\end{figure*} 
The  measured $\Dphi$ modulation of the $P_2^{\rm (CI)}$ correlation function and its dependence on collision centrality is also of particular interest. One finds, as shown in Fig.~\ref{fig:cP2CI}, that the $P_2^{\rm (CI)}$ correlator measured in most central \PbPb\ collisions exhibits an away-side double ridge or hump structure that extends across the full $\Delta \eta$ acceptance. This implies the presence of a very strong $v_3(P_2^{\rm (CI)})$ component relative to the $n=2$ component in that collision centrality bin. This and the observed evolution of the Fourier decompositions of the 
 $P_2^{\rm (CI)}$ correlator, compared to expectations based on a simple flow ansatz, in fact lend further support 
 to the notion that the observed $\Delta \eta$ correlations are evidence for collective anisotropic flow relative to the 
 collision reaction plane~\cite{Adam:2017ucq}. It is interesting to note, however, that the three models produce a fairly flat away side vs. $\Delta\varphi$, even a small dip at $\Delta\varphi=\pi$, in most central collisions in Fig.~\ref{fig:cDphiCI}.  Remarkably, the depth of the dip  obtained with UrQMD is the strongest although this model produces  a rather poor $\Delta \eta$ dependence representation of the two particle correlation data. It is indeed not the presence of the dip that constitute evidence for collectivity but its near invariance with $\Delta \eta$ and the quantitative agreement between 
 the observed magnitude of that ($v_{3}$) harmonic component in $P_2^{\rm (CI)}$ relative to the flow ansatz. Such (away) $\Delta \eta$  invariance of the $\Dphi$ modulation is qualitatively reproduced by both the AMPT and EPOS models  but these models require further tuning to perfectly match the   $R_2^{\rm (CI)}$ and $P_2^{\rm (CI)}$ correlation functions reported by the ALICE collaboration.

\subsection{Charge Dependent (CD) Correlation Functions} 
\label{sec:ResultsCD}

\begin{figure*}[htb] 
\includegraphics[scale=0.80,clip=true,trim=0pt 290pt 0pt 300pt]{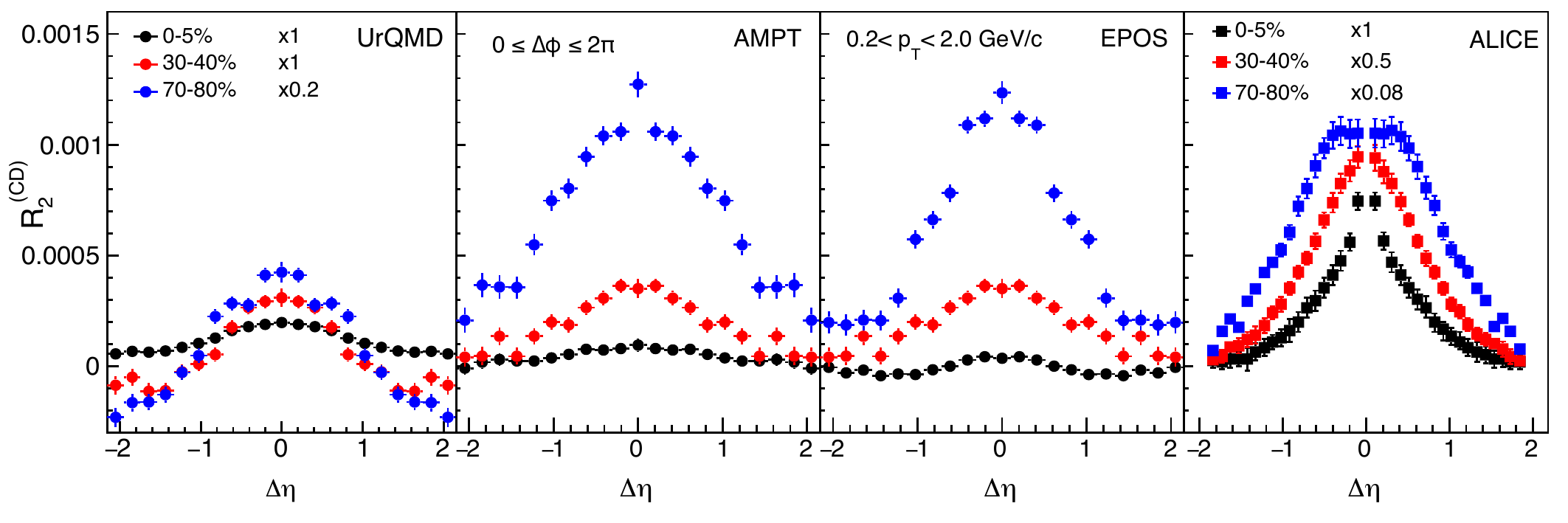}
\caption{Projections of $\rm{R_2^{(CD)}}$ correlators of charged hadrons obtained with UrQMD, AMPT and EPOS event generators compared to projections of the correlators measured by the ALICE collaboration~\cite{PhysRevC.100.044903} in \PbPb\ collisions at $\snn$ = 2.76 TeV shown in Figs.~\ref{fig:cR2CD} and \ref{fig:cP2CD}. Scaling factors listed in the left panel apply to the three model calculations.}
\label{fig:cR2CD_deta}
\end{figure*}

\begin{figure}[ht!]
\begin{center} 
\includegraphics[scale=0.4]{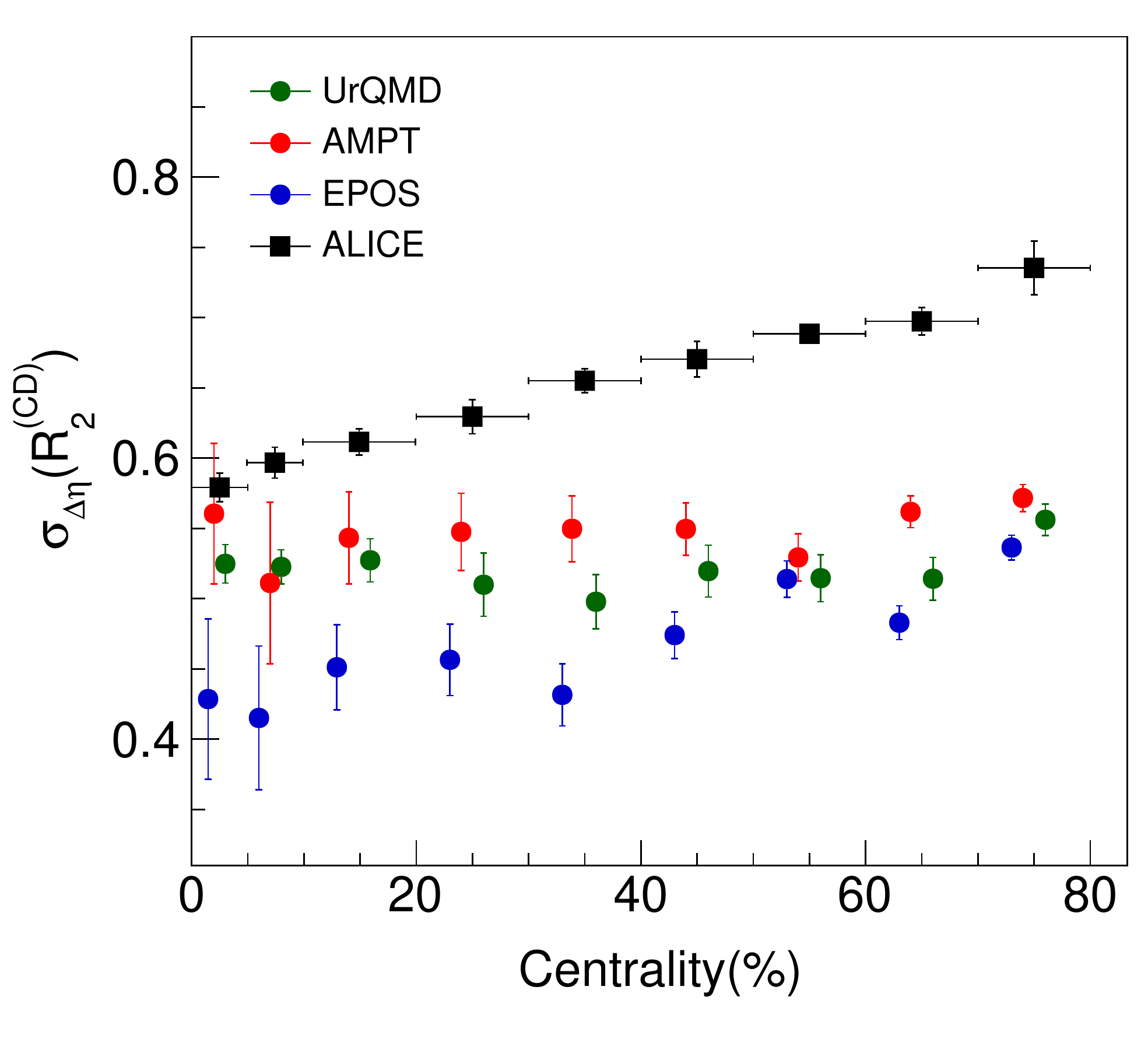}
\caption{Longitudinal rms width ($\sigma_{\Delta\eta}$) of the near-side  peak of  $\rm{R_2^{(CD)}}$ correlators (shown in Fig.~\ref{fig:cR2CD}) plotted as a function of collision centrality. Widths computed with based on the correlators computed with UrQMD, AMPT and EPOS are compared to rms values reported by the ALICE collaboration~\cite{PhysRevC.100.044903}.}
\label{fig:cR2CD_detaRmsWidth}
\end{center} 
\end{figure}

Figures~\ref{fig:cR2CD} -- \ref{fig:cP2CD}  present  comparisons  of UrQMD, AMPT, and EPOS calculations with ALICE measurements of the $R_2^{\rm (CD)}$ and $P_2^{\rm (CD)}$ correlators, respectively.  Projections of the $R_2^{\rm (CD)}$  correlators  onto $\Delta\eta$ are shown in  
Fig.~\ref{fig:cR2CD_deta}.  The calculated $P_2^{\rm (CD)}$ model correlators shown in Fig.~\ref{fig:cP2CD}  have small amplitudes and rather limited statistical accuracy. Their projections are thus of limited interest and are not shown in this paper.
We first remark that all  three models qualitatively reproduce the presence of the prominent near-side peak of the   $R_2^{\rm (CD)}$  correlator. Note, however, that the broad dip centered at $(\Delta\eta,\Delta\varphi)=(0,0)$ observed in data is largely associated with Hanbury - Brown -- Twist (HBT) correlations and
is thus not expected to be reproduced by the model simulations discussed in this work given they do not feature an HBT afterburner. All three models also produce an away-side tail in most peripheral collisions. This tail is largely caused by the decay of  resonances. For instance, decays of low-$p_{\rm T}$ $\rho^0$-mesons   yield nearly back-to-back pions with small $\Delta\eta$ pair separation.  The models also qualitatively reproduce the progressive suppression of this tail in more central collisions owing to an increase of the produced parent particles average transverse momentum $\la p_{\rm T}\ra$. However, the models produce near-side peak amplitudes and  collision centrality evolution that are somewhat inconsistent with those observed experimentally. As shown in Fig.~\ref{fig:cR2CD_detaRmsWidth}, they also poorly reproduce the magnitude and collision centrality evolution of the longitudinal rms width of the near-side peak of the $R_2^{\rm (CD)}$ correlator. The measured
rms widths (black squares) exhibit a distinct narrowing, approximately 30\%, with increasing collision centrality whereas
AMPT and UrQMD produce peak rms widths that are independent, within statistical uncertainties, of the collision centrality. The rms $\sigma_{\Delta\eta}$ is calculated according to $\sigma_{\Delta\eta}^2=\sum_{i} \left(R_2^{({\rm CD})}(\Delta\eta_i)-P\right) \Delta\eta_i$, where
the sum is taken across all $\Delta\eta_i$ bins and $P$ represents the correlation pedestal (minimum) evaluated at $|\Delta\eta|=2$. 
  EPOS produces a narrowing of the near-side peak but the widths it produces are too narrow by approximately 30\%. The excessive narrowness of the peak likely results from the dominance of corona particles in this EPOS calculation of  $R_2^{\rm (CD)}$.  Indeed, the fact that the core component likely underestimate the correlator strength given it does not fully implement event-by-event charge conservation implies the  correlator is dominated by corona particles. Given the average radial flow imparted to  corona particles is much larger than the average (core), one then observes an excessive kinematic narrowing of the near-side peak.  In the case of UrQMD, the weak amplitude of the near-side peak may be in part due to an insufficient number of ``high-mass" resonances. The weakness of the peak  observed in AMPT and EPOS calculations, however, is most likely due to their incomplete handling of charge-conservation. 

Shifting our attention to the $P_2^{\rm (CD)}$ correlator calculations shown in Fig.~\ref{fig:cP2CD}, we first note that the model calculations and ALICE data are considerably  challenged by the rather weak magnitude of the $\la\Delta p_{\rm T}\Delta p_{\rm T}\ra$ correlator.  We note, nonetheless, that UrQMD and AMPT both produce a narrow near-side peak in central collision, albeit with too weak an amplitude
relative to the near-side peak observed in the data. By contrast, EPOS produces a narrow valley in lieu of a peak. A negative value of the $\la\Delta p_{\rm T}\Delta p_{\rm T}\ra$ correlator is indicative of the dominance of correlation between low and high-$p_{\rm T}$ particles (i.e., one particle below and one particle above the mean  $\la p_{\rm T}\ra$. By contrast, the ALICE data feature a positive $\la\Delta p_{\rm T}\Delta p_{\rm T}\ra$  correlator, which indicates that correlations are dominated by correlation of particle pairs involving particles  that are both below or above  $\la p_{\rm T}\ra$. Clearly, all three models require considerable tuning before they can reproduce $R_2^{\rm (CD)}$ and $P_2^{\rm (CD)}$ correlators reported by the ALICE collaboration.

\section{Summary }\label{sec:summary}

We presented comparisons  of  calculations with the UrQMD, AMPT, and EPOS models of two-particle differential number correlators, $R_2$,  and transverse momentum correlators, $P_2$,  with data  recently reported  by the ALICE collaboration. 
We find that while  these models  can approximately reproduce the evolution of the strength of these correlators they cannot  satisfactorily reproduce the the detailed shape and features of the measured like-sign (LS), unlike-sign (US), charge independent (CI), and charge dependent (CD) correlation functions, and their collision centrality evolution.  UrQMD is arguably challenged the most given it is unable to reproduce the strong $\Delta\varphi$ modulation and  the nearly $\Delta\eta$ invariant correlation strength observed with the  $R_2^{\rm (CI)}$ and  $P_2^{\rm (CI)}$ correlators. It also  underestimates the magnitude of the near-side peak of the measured $R_2^{\rm (CD)}$ and  $P_2^{\rm (CD)}$ correlators.  AMPT produces a qualitatively better description of the data  given that it produces sizable flow-like modulations in $R_2^{\rm (CI)}$  and  $P_2^{\rm (CI)}$. However, it also  underestimates the magnitude of 
the near-side peak of   $R_2^{\rm (CD)}$ and  $P_2^{\rm (CD)}$ correlators, as a result most likely of improper handling of charge conservation. EPOS produces a relatively good match to the data: It qualitatively reproduces  the shape, strength, and collision centrality evolution of the $R_2^{\rm (CI)}$  and  $P_2^{\rm (CI)}$ correlators. It also produces a sizable near-side peak in $R_2^{\rm (CD)}$.  However, irrespective of the fact that it does not feature an HBT afterburner, it is unable to reproduce the magnitude of this correlator's near-side  and its narrowing from peripheral to central collisions. Oddly, it also produces a sizable correlation dip centered at $(\Delta\eta,\Delta\varphi)=(0,0)$ in $P_2^{\rm (CD)}$ for 0-5\% most central collisions,
in drastic contrast to the  peak observed experimentally. Given the structure of the $P_2$ correlator, this suggests that EPOS overemphasizes correlations between low $p_{\rm T}$ (i.e., $p_{\rm T} < \la p_{\rm T}\ra$) and high $p_{\rm T}$  (i.e., $p_{\rm T} > \la p_{\rm T}\ra$) particle pairs. It is noteworthy that through its corona component, EPOS is able to reproduce a sizable fraction of the observed near-side peak of $R_2^{\rm (CD)}$, although its core component is not expected to yield a significant charge dependent correlation strength given the Cooper-Frye mechanism used in EPOS for hadronization of the hydrodynamics core does not necessarily conserve charge locally on an event-by-event basis. 

The AMPT and EPOS models have had great successes in reproducing single particle $p_{\rm T}$ spectra, ratios of particle abundances and their collision centrality evolution, as well as the magnitude of measured flow coefficients. In this study, we find that the $\Delta\varphi$ modulations produced by AMPT best match the measured coefficients, while EPOS tend to slightly overestimate their magnitude. As such, it is clear that both models capture much of the production and transport dynamics in \PbPb\ collisions at LHC. Yet, they do not properly reproduce the key features  of the measured $R_2^{\rm (CI,CD)}$  and  $P_2^{\rm (CI,CD)}$ correlators. This most likely stems from  
a poor handling, on an event-by-event basis, of charge, strangeness, and baryon number conservation. This is rather unfortunate given that measurements of CD correlations, or equivalently measurements of balance functions,  potentially have the capacity to inform us about the production time of up, down, and strange quarks in AA collisions. 
Are there two stages  of quark production as postulated in Ref.~\cite{S.PrattPRL:2000BalFun1st}? Does baryon production and conservation~\cite{Acharya:2019izy} play a role during the early stages of collision systems evolution, or is the production of baryon anti-baryon pairs solely a stochastic process
taking place during the hadronization stage of the QGP?

We have  shown that  the $R_2^{\rm (CI,CD)}$  and  $P_2^{\rm (CI,CD)}$ correlators are quite sensitive  to the details particle production dynamics and more specifically  model implementations of charge, strange and baryon conservation. 
Given CI, CD correlators, and balance functions are 
in principle sensitive to the viscosity and the diffusivity of the matter produced in A--A collisions~\cite{Pratt:2019pnd}, 
further development of  theoretical models is required to account for charge, strange, and baryon conservation so that
observables such as those discussed in this paper can be used to further our understanding of the properties of the matter produced in A--A collisions and most particularly the QGP.  We stress that inclusion of local quantum number conservation in a modified Cooper-Frye formula, in particular, would enable considerable 
advances in the interpretation of published ALICE results~\cite{Adam:2017ucq,PhysRevC.100.044903} while techniques to properly implement
charged, strange, and baryon currents  ab-initio in hydrodynamics are fully developed. 

In closing, we note that significant advances  are being made towards the implementation of local quantum number conservation  based on  Metropolis sampling  developed  towards the particlization of  fluid cells in hydrodynamic simulation of the evolution of A--A collisions\cite{Oliinychenko:2019zfk,Oliinychenko:2020cmr,Vovchenko:2020kwg}. As these methods locally preserves the  conservation of energy, momentum, baryon number, strangeness, and electric charge microcanonically, they should enable significant advances in studies of two- and multiple particle differential correlators. Future  studies shall thus examine the impact of the deployment of these and similar methods on integral and differential correlation functions~ \cite{Schwarz:2017bdg,Braun-Munzinger:2019yxj}.


\newenvironment{acknowledgement}{\relax}{\relax}
\begin{acknowledgement}
\section*{Acknowledgements}
SB and CP wish to thank Klaus  Werner for fruitful discussions. 
This work was supported in part by the United States Department  of Energy, Office of Nuclear Physics (DOE NP), United States of America under Grant No. DE-FOA-0001664 and the U.S. Department of Energy Office of Science under contracts DE-SC0013391 and DE-SC0015636.
 SB also acknowledge the support of the Swedish Research Council (VR).
The authors acknowledge the Texas Advanced Computing Center (TACC) at the University of Texas at Austin for
providing computing resources that have contributed to the research results reported within this paper. URL: http://www.tacc.utexas.edu. The authors also thank the GSI Helmholtzzentrum f\"ur Schwerionenforschung for providing the computational resources needed for producing the UrQMD events used in this analysis. 
\end{acknowledgement}

\bibliographystyle{apsrev4-1}
\bibliography{Model_Corr_R2P2}
\end{document}